

%
%

\magnification=\magstephalf
\def\omit#1{}

\def\comm#1{}

\def\textskip{\baselineskip = 12pt plus 1pt minus 1pt}
\def\capskip{\baselineskip = 10 pt}
\parskip=6pt

\font\csc=cmcsc10   
\font\bfcbigg=cmssdc10 scaled \magstep2  
\def\dirac{{\scriptscriptstyle\rm DIRAC}}
\def\etal{{\it et~al.}}
\def\eg{{\it e.g.~}}
\def\ie{{\it i.e.~}}
\def\cf{{\it c.f.~}}
\def\etc.{{\it etc.~}}
\def\vct#1{\skew{-2}\vec#1} 
\def\kms{{\rm\,km\,s^{-1}}}
\def\mpc{{\rm\,Mpc}}
\def\hmpc{{h^{-1}\rm\,Mpc}}
\def\frac#1/#2{\leavevmode\kern.1em\raise.6ex\hbox{\the\scriptfont0 #1} 
       \kern-.15em/\kern-.2em\lower.7ex\hbox{\the\scriptfont0 #2}} 

\newbox\grsign \setbox\grsign=\hbox{$>$} \newdimen\grdimen \grdimen=\ht\grsign
\newbox\simlessbox \newbox\simgreatbox
\setbox\simgreatbox=\hbox{\raise.5ex\hbox{$>$}\llap
     {\lower.5ex\hbox{$\sim$}}}\ht1=\grdimen\dp1=0pt
\setbox\simlessbox=\hbox{\raise.5ex\hbox{$<$}\llap
     {\lower.5ex\hbox{$\sim$}}}\ht2=\grdimen\dp2=0pt

\def\simlt{\mathrel{\copy\simlessbox}}

\def\chaphead{}
\newcount\eqnumber
\eqnumber=1
\def\new{{\rm\chaphead\the\eqnumber}\global\advance\eqnumber by 1}
\def\eqref#1{\advance\eqnumber by -#1 \chaphead\the\eqnumber
     \advance\eqnumber by #1 }
\def\last{\advance\eqnumber by -1 {\rm\chaphead\the\eqnumber}\advance
     \eqnumber by 1}

\def\dt{\partial_t}

\def\del{\delta}
\def\grad{{\vct\nabla}_x}
\def\div{{\vct\nabla}_x\cdot}

\def\gradsq{\nabla^2_x}
\def\posx{(\vct x, t)}
\def\posy{(\vct y, t)}

\def\denvel{{density-to-velocity\ }}
\def\velden{{velocity-to-density\ }}
\def\undertext#1{$\underline{\smash{\hbox{#1}}}$}

   \font\eightrm=cmr8   \font\sixrm=cmr6
   \font\eighti=cmmi8   \font\sixi=cmmi6
  \font\eightsy=cmsy8  \font\sixsy=cmsy6
  \font\eightbf=cmbx8  \font\sixbf=cmbx6
  \font\eighttt=cmtt8
  \font\eightit=cmti8
  \font\eightsl=cmsl8

\catcode`@=11
\newskip\ttglue

\def\eightpoint{\def\rm{\fam0\eightrm}
  \textfont0=\eightrm \scriptfont0=\sixrm \scriptscriptfont0=\fiverm
  \textfont1=\eighti  \scriptfont1=\sixi  \scriptscriptfont1=\fivei
  \textfont2=\eightsy \scriptfont2=\sixsy \scriptscriptfont2=\fivesy
  \textfont3=\tenex \scriptfont3=\tenex \scriptscriptfont3=\tenex
  \textfont\itfam=\eightit  \def\it{\fam\itfam\eightit}%
  \textfont\slfam=\eightsl  \def\sl{\fam\slfam\eightsl}%
  \textfont\ttfam=\eighttt  \def\tt{\fam\ttfam\eighttt}%
  \textfont\bffam=\eightbf  \scriptfont\bffam\sixbf
   \scriptscriptfont\bffam=\fivebf \def\bf{\fam\bffam\eightbf}%
  \tt \ttglue=0.5em plus .25em minus.15em
  \setbox\strutbox=\hbox{\vrule height7pt depth2pt width0pt}%
  \let\sc=\sixrm  \let\big=\eightbig  \normalbaselines\rm}


\centerline{\bfcbigg TESTING THE GRAVITATIONAL INSTABILITY HYPOTHESIS?}
\bigskip
\centerline{\csc Arif Babul$^1$, David H. Weinberg$^2$,
Avishai Dekel$^{3}$ \& Jeremiah P. Ostriker$^4$}
\bigskip
\line{\hfill\vbox{\hbox{\it $^1$ Canadian Institute for Theoretical
Astrophysics, Toronto, Ontario M5S 1A1, Canada}
\hbox{\it $^2$ Institute for Advanced Study, Princeton, NJ 08540, USA}
\hbox{\it $^3$ Racah Institute of Physics, Hebrew University, Jerusalem,
Israel}
\hbox{\it $^4$ Princeton University Observatory, Princeton, NJ 08544, USA}
\medskip
\hbox{E-mail: babul@cita.utoronto.ca, dhw@guinness.ias.edu,
dekel@venus.huji.ac.il,}
\hbox{\phantom{E-mail: }jpo@astro.princeton.edu}}
\hfill}
\bigskip
\medskip
\noindent
{\bf NOTE:} {\it A postscript version of this paper, with embedded figures, is
available by anonymous ftp.  Contact dhw@guinness.ias.edu for details.}
\bigskip
\medskip
\centerline{\bf ABSTRACT}
\bigskip

We challenge a widely accepted assumption of observational
cosmology: that successful reconstruction of observed galaxy density
fields from measured galaxy velocity fields (or vice versa), using
the methods of gravitational instability theory,
implies
that the
observed large-scale structures and large-scale flows were produced
by the action of gravity.  This assumption is false,
in that there exist non-gravitational theories that pass the reconstruction
tests and gravitational theories with certain forms of biased galaxy
formation that fail them.
Gravitational instability theory predicts specific correlations between
large-scale velocity and mass density fields, but the same correlations arise
in any model where (a) structures
in the galaxy distribution grow from homogeneous initial conditions
in a way that satisfies the continuity
equation, and (b) the present-day velocity field is irrotational and
proportional to the
time-averaged velocity field.  We demonstrate these assertions using
analytical arguments and $N$-body simulations.

If large-scale structure formed by gravitational instability, then the ratio of
the galaxy density contrast to the divergence of the velocity field
yields an estimate of the density parameter $\Omega$ (or, more generally,
an estimate of $\beta\equiv\Omega^{0.6}/b$, where $b$ is an assumed constant
of proportionality between galaxy and mass density fluctuations).
In non-gravitational scenarios, the values of $\Omega$ or $\beta$ estimated
in this way may
fail to
represent the true cosmological values.
However, even if non-gravitational forces initiate and shape the growth
of structure, gravitationally induced accelerations can dominate the
velocity field at late times, long after the action of any non-gravitational
impulses.  The estimated $\beta$ approaches the true value in such cases,
and in our numerical simulations the estimated
$\beta$ values are reasonably accurate for both gravitational and
non-gravitational models.

Reconstruction tests that show
correlations between galaxy
density and velocity fields can rule out some physically interesting
models of large-scale structure.  In particular, successful reconstructions
constrain the nature of any bias between the galaxy and mass distributions,
since processes that modulate the efficiency of galaxy formation on large
scales in a way that violates the continuity equation also produce a
mismatch between the observed galaxy density and the density inferred
from the peculiar velocity field.  We obtain successful reconstructions
for a gravitational model with peaks biasing, but we also show
examples of gravitational
and non-gravitational models that fail reconstruction tests because of
more complicated modulations of galaxy formation.

\medskip
\line{{\it Subject Headings:} cosmology: theory --- large-scale structure of
the Universe --- galaxies: clustering\hfil}
\bigskip
\bigskip

\vfill\eject
\textskip

\centerline{\bf 1. Introduction }
\bigskip

The fundamental paradigm of modern cosmology postulates a dynamic universe
in which the matter distribution, nearly homogeneous at early times, is
cast into motion and consequently evolves into a network
of swells and troughs corresponding to converging and diverging flows.
The general belief is that the motions have been produced by the force
of gravity.
But do we have direct empirical evidence for this belief?
The standard methods for
analyzing large-scale velocity and density fields {\it assume}
that structure formed by gravitational instability, and they have sometimes
been thought to {\it test} this assumption as well.
Here we show that these methods test a rather different combination of
assumptions, the most fundamental of which is simply the continuity
equation, and that they do not test gravity {\it per se}.
We demonstrate this point
first by considering the relation between
velocity and density fields for a broad class of models in the linear
approximation, and then by analyzing numerical simulations of
a gravitational model
and a specific non-gravitational model, the explosion
scenario developed by Ostriker \& Cowie (1981), Ikeuchi (1981), and others.
Throughout this paper, we use explosions only as a convenient
example of a non-gravitational model of structure formation; our arguments
and results apply to a variety of non-gravitational scenarios.

In the gravitational instability picture of structure formation,
gravitational forces amplify small fluctuations, present in the early
universe, into the galaxies, clusters, voids, and superclusters observed today.
One might think of testing this hypothesis
by searching for the expected correlations
between large-scale density fluctuations and peculiar velocity flows.
``Tests'' of this sort take one of two forms.  In the \denvel approach,
one starts with the observed galaxy density field, assumes some relationship
between the galaxy distribution and the underlying mass distribution,
predicts the peculiar velocities, and compares them to observations.  In the
\velden approach, one starts with the peculiar velocities, predicts the
mass density field, and compares it to the galaxy density field.

The first attempts at \denvel analyses were spurred by the discovery of
the dipole anisotropy in the cosmic microwave background (CMB) (Smoot,
Gorenstein \& Muller 1977) and its interpretation as the reflex of the
peculiar motion of the Local Group (LG) (\eg Smoot \& Lubin 1979;
Lubin \& Villela 1986).  There have been a number of attempts to account for
the origin of this motion in terms of the gravitational acceleration
produced by the inhomogeneous matter distribution around the LG.  Some
of these studies compare the CMB dipole to the dipole of the distribution
of optical or IRAS galaxies on the sky (\eg Meiksin \& Davis 1986; Lahav 1987;
Harmon, Lahav \& Meurs 1987; Lahav, Rowan-Robinson \& Lynden-Bell 1988;
Lynden-Bell, Lahav \& Burstein 1989).  More recent studies have utilized the
redshift surveys
of IRAS galaxies (Rowan-Robinson \etal\ 1990; Strauss \etal\ 1992a) and of rich
galaxy clusters (Scaramella \etal\ 1991).  The most important recent
advances in \denvel analyses rely on
systematic surveys of galaxy peculiar velocities,
which use the Tully-Fisher or $D_n-\sigma$ relations to obtain
redshift-independent distance estimates (\eg Aaronson \etal\ 1986;
Dressler \etal\ 1987; see review by Burstein 1990).
With these data and a large, homogeneously selected redshift survey, one
can attempt to predict peculiar velocities of a large number of galaxies
instead of the LG alone (Strauss \& Davis 1991; Kaiser \etal\ 1991;
Davis \etal, in preparation).

Peculiar velocity surveys also make it possible to reverse the
order of analysis and predict the mass density field starting
from the observed motions.  There are two methods for doing so.  In the first
method, dubbed POTENT, a 3-dimensional velocity field is constructed from the
radial velocity data, under the assumption that the velocity field is a
potential flow (Bertschinger \& Dekel 1989;
Dekel, Bertschinger \& Faber 1990, hereafter DBF).
The resulting velocity field can be transformed into a density field.
In the second method, developed by Kaiser \&
Stebbins (1991), the 3-d velocity field (and therefore the density field) is
represented by a set of Fourier modes, which are iteratively relaxed until
the most probable configuration, given the radial velocity data, is achieved.
Recently, Dekel \etal\ (1993) have compared
the POTENT mass density field recovered
from peculiar velocity data to the galaxy density field of the
1.936 Jy IRAS redshift survey
(Strauss \etal\ 1992b).  A follow-up
comparison of more extensive velocity data to the 1.2 Jy redshift survey
(Fisher 1992) is underway.
Hudson and Dekel (1993) have performed a similar comparison
between the POTENT mass density field and the density of optical galaxies.

Results of the multitude of dipole, density-to-velocity,
and \velden studies have been
varied, and we will not attempt any detailed comparisons to observations
in this paper.  It is probably fair to say that, in most of the
studies, the observations are consistent with the predictions of gravitational
instability theory to within the rather large
random and systematic errors in the observational data and
the theoretical approximations.  Redshift and distance
data are accumulating rapidly, so the observational situation should improve
significantly over the next few years.

The theoretical underpinning of studies comparing velocity and density fields
is usually couched in the language of the gravitational instability hypothesis.
Within this framework, the relative amplitudes of density fluctuations and
peculiar velocities can be used to estimate the value of the cosmological
density parameter $\Omega$ (or, more generally, the value of
$\beta\equiv\Omega^{0.6}/b$, where the
``bias'' factor $b$ is an assumed constant of proportionality between
fluctuations in the galaxy density and the underlying mass fluctuations,
see equation [1] below.)
However, if the velocities are not gravitationally induced, then the
derived value of $\beta$ loses its fundamental cosmological significance,
since it need
not reflect the true value of the density parameter.
It is therefore important to ask whether observing the predicted
correspondences between velocity and density fields
{\it verifies} the gravitational instability hypothesis itself
(as sometimes implied in the literature, \eg Yahil 1988).
Because the correspondences are usually derived from the equations of
gravitational instability, the answer would appear to be ``yes.''
However, if the primordial mass distribution is homogeneous, then converging
flows will produce overdense regions and diverging flows will produce
underdense regions, whether or not the peculiar velocities
arise from gravitational accelerations.
As a result, agreement between predicted and observed density and velocity
fields {\it does not} necessarily argue against
theories in which large-scale structure and peculiar velocities are generated
primarily by non-gravitational forces such as cosmological explosions
(Ostriker \& Cowie 1981; Ikeuchi 1981; Ostriker, Thompson \& Witten 1986)
or radiation pressure instabilities (Hogan \& Kaiser 1983; Hogan \& White
1986).

A brief consideration of the equations governing linear velocity and density
fields illustrates this point more directly.
In the linear theory of gravitational instability, peculiar velocity is
proportional to the gradient of the gravitational potential, $\vec v \propto
-\vec\nabla\Phi_g$, and the gravitational potential is related to the density
fluctuations $\delta$ through the cosmological generalization of Poisson's
equation.  Consequently, $\vec\nabla\cdot\vec v \propto -\nabla^2\Phi_g
\propto -\delta$. However, if the linearized continuity equation holds
($\div\vct{v}\propto -\partial\delta/\partial t)$,
and the initial galaxy distribution
is homogeneous, and the final velocities are proportional to their
time-averaged values, then $\div \vct{v} \propto -\delta$
{\it regardless of the velocities' physical origin.}
Therefore, one would expect
that the predicted velocity or density fields will match observations under
conditions more general than those
of the standard gravitational instability picture.

Conversely, one can
anticipate that even if the gravitational instability model is correct,
it may fail the density-velocity tests if galaxy formation has been
modulated by a
non-gravitational process that violates the continuity equation.
``Biased'' galaxy formation
is introduced specifically to allow
for such effects in gravitational models (see review by Dekel \& Rees 1987).
If the efficiency of galaxy formation is enhanced in some environments and
suppressed in others, then the current distribution of galaxies cannot be
obtained by taking an initially uniform distribution of galaxy markers and
moving them under gravitational forces to their positions at $z=0$.
A linear, scale-independent bias model is often adopted for simplicity,
$$
(\delta N/N)_{gal} =
b(\delta M/M).
\eqno(\new)
$$
In this case the galaxy density contrast $(\delta N/N)$ is proportional
to the mass density contrast $(\delta M/M)$,
which does obey the continuity equation.
However, linear bias is at best a theoretical convenience, and the
situation in realistic gravitational instability models is more complex
(see, {\it e.g.}, the detailed numerical study of Cen \& Ostriker 1992).
We will return to this theme later in the paper.

In the next section, we present the analytic theory of velocity and density
fields in greater detail, focusing on the linear regime.
In the sections that follow, we apply the usual \denvel and \velden tests
to $N$-body simulations of the explosion scenario and of gravitational
models.
The simulations are described in \S 3.
In \S 4 we apply the tests to these simulations, considering both unbiased
models, which necessarily obey the continuity equation, and simply biased
versions of the gravitational models, in which the galaxy density contrast
is closely related to the mass density contrast.
In \S 5 we discuss models that fail the density-velocity tests, with
particular attention to models in which
non-trivial biasing schemes make
the evolution of the galaxy distribution
violate the continuity equation.
In \S 6, we apply the POTENT method to some of our simulations to
demonstrate that
it successfully recovers the 3-d velocity field
and the density field from radial velocities alone,
regardless of whether the proper velocities are induced
by gravity, so long as the
continuity equation is satisfied and the velocity field is
irrotational and proportional to its time-average.
We discuss the implications of our results in \S 7.

\bigskip
\centerline{\bf 2. Theory}
\medskip

In the standard gravitational instability model, the equations governing
the mass distribution are (Peebles 1980):\hfil\break
\line{the continuity equation,\hfill}
$$\dt\del + \div\left[(1+\del)\vct{v}\right] =0, \eqno(\new)$$
\line{the Euler equation,\hfill}
$$
\dt\vct{v} + 2H(t)\vct{v} + \left(\vct{v}\cdot\grad\right)\vct{v}=
   - \grad \Phi_g, \eqno(\new)
$$
\line{and the Poisson equation,\hfill}
$$\gradsq\Phi_g = 4\pi G \bar{\rho}(t)\del
= {3\over 2} H^2(t) \Omega(t) \del.  \eqno(\new)$$
In these equations,
$\del\posx\equiv [\rho\posx - \bar{\rho}(t)] /\bar{\rho}(t)$
is the density contrast,
and $\vec{x}$ and $\vec{v} \equiv \dot{\vec{x}}$ are the position and the
peculiar velocity, respectively, in comoving units
The corresponding physical quantities are $a\vec{x}$ and $a\dot{\vec{x}}$,
where $a(t)$ is the expansion factor of the universe.
Also, $\dt\delta \equiv (\partial\delta /\partial t)$,
$\dt\vct{v} \equiv (\partial\vct{v} /\partial t)$,
$H(t)\equiv\dot a/ a$ is the Hubble parameter,
$\Omega(t) \equiv 8\pi G\bar\rho(t)/3H^2(t)$ is the density parameter,
and $\Phi_g$ is the comoving gravitational potential (the corresponding
physical quantity being $a^2\Phi_g$).

Together with the boundary conditions $\del \rightarrow 0$ and
$\vct{v} \rightarrow 0$ as $t \rightarrow 0$, equations
(\eqref{3})--(\eqref{1}) may be taken as a definition of the standard
gravitational instability model.  They can be solved by linear perturbation
theory in the limit of small fluctuations.  Of particular interest
to us is the linear, growing-mode relation between the velocity and
density fields:
$$ \eqalignno{
\div\vct{v}\posx &= -Hf(\Omega)\del\posx , &(\new{\hbox{a}}) \cr
\vct{v}\posx     &= {Hf(\Omega)\over 4\pi} \int d\vct y\ \del\posy
{(\vct y-\vct x) \over\left\vert \vct y -\vct x\right\vert^3} ,
&(\eqref{1}{\hbox{b}})
}
$$
with $f(\Omega) \approx \Omega^{0.6}$ (Peebles 1980).
Equation (\eqref{1}b) follows from (\eqref{1}a) because the growing-mode
velocity field is irrotational (vorticity-free), and it can therefore
be derived from a velocity potential, $\vec{v}=-\grad\Phi_v$.

Equations (\eqref{1}a) and (\eqref{1}b) are the basic equations for
\velden and \denvel reconstructions, respectively, in the linear regime.
Such reconstructions directly test a combination
of two hypotheses: (1) that $\div\vct{v} \propto -\delta$, and
(2) that the velocity field is irrotational.
In a \denvel reconstruction, the irrotational
assumption is used to compute the full velocity field
from its divergence.  A \velden reconstruction that utilized the
3-dimensional velocity field would not require this assumption, but
observations provide only the radial component of the velocity field,
and one must assume a potential flow in order to build the full, 3-dimensional
field.  If structure formed by gravitational instability, then the
constant of proportionality between $\div\vct{v}$ and $-\del$
provides a measure of $f(\Omega)$; by working in velocity units, one
can avoid any dependence on the Hubble constant $H$.

It has often been argued that succesful \velden and \denvel reconstructions
provide confirmation of
the gravitational instability scenario (\eg Yahil 1988; similar arguments
appear in much of the literature on cosmological dipoles and velocity fields).
However, the two conditions tested by these methods are by no means
unique to gravitational instability models.
Whatever the source of peculiar velocities,
the linearized form of the continuity equation (2) implies that
$$\dt\del = -\div\vct{v} . \eqno(\new)$$
If the galaxy distribution is smooth at high redshift, one
can integrate both sides of this equation from $t=0$ (and $\delta=0$)
to $t=t_0$ to obtain
$$\delta = -t_0\,\div\langle\vct{v}\rangle_t , \eqno(\new)$$
where $\langle\vct{v}\rangle_t \equiv \int_0^{t_0} \vct{v} dt/t_0$ is the
time-averaged velocity field.
Therefore, any model that has homogeneous initial conditions and
obeys the continuity equation will yield successful reconstructions,
provided that its
present-day velocity field is irrotational and is proportional to
the time-averaged velocity field.

In fact we require only that the
present-day velocity field be proportional to the irrotational part of the
time-averaged field, because a divergence-free component that might have
existed in the past but decayed since would not leave behind any
associated density perturbations. With this qualification, the proportionality
condition is necessary as well as sufficient; an initially homogeneous
model that obeys the continuity equation and satisfies
$\div\vct{v} \propto -\delta$ today must, by equation (\eqref{1}),
have a present-day velocity divergence that is proportional to
its time-averaged value.
Irrotationality does not follow from the continuity equation alone, however;
it must be adopted as a separate condition.
Expansion of the universe always erodes vorticity,
so an irrotational velocity field is likely to arise
in many scenarios, both gravitational and non-gravitational,
at least on scales in the linear regime.
Irrotationality is not in itself a sufficient condition
for succesful reconstructions, since it does not guarantee proportionality
between the present and time-averaged velocity fields.

In the sections that follow, we will present several examples of
non-gravitational models that pass the usual density-velocity tests.
We start by analyzing the behavior of some idealized models in the
linear regime, then proceed in \S 4 to numerical simulations of more
realistic, non-linear models.

\bigskip
\centerline{\bf 2.1 A Simple Example: A One-component Universe with No Gravity
($\Omega=0$)}
\medskip

As our first example, we consider the simple but extreme case
of an $\Omega=0$ universe filled with massless test particles.
In such a universe, there are no gravitational forces.
We assume that the test particles are uniformly distributed at very early
times,
and that they acquire peculiar velocities from non-gravitational
forces that produce an impulsive perturbation,
$\vct{V}(\vct x) \delta^\dirac(t-t_e)$, acting at time $t=t_e$.
Apart from this initial impulse, we
ignore any pressure forces.

On scales where density
inhomogeneities and the associated velocities are small, the
governing equations are the linearized continuity equation (\eqref{2})
and the linearized Euler equation,
$$\dt\vct{v}+2H(t)\vct{v} =\vct{V}(\vct x)
\delta^\dirac(t-t_e).
\eqno(\new)$$
By combining these, we find that for $t>t_e$ the density
perturbations  evolve according to
$$ \dt^2\del+2H(t)\dt\del= 0, \eqno(\new)$$
where the effects of the initial perturbations are absorbed into the initial
conditions at time $t=t_e$:
$\delta(\vct{x},t_e)=0$, and
$\dt\delta(\vct{x},t_e)=\zeta (\vct x) \equiv -\div\vct{V}(\vct x)$.
Noting that $H(t)=1/t$ and $a=t/t_e$ (the expansion scale factor is normalized
so that $a_e=1$), we find that
$$\eqalign{ \delta\posx=& \zeta(\vct x)t_e (a-1)/a, \cr
t\dt\delta\posx=&\zeta(\vct x)t_e/a.  \cr}\eqno(\new)$$
At any finite time, $\dt\del$ and $\del$ are clearly
{\it proportional} to each other,
\ie the ratio $\del/\dt\del$ is independent of spatial position.

We can combine equations (\eqref{5}) and
(\eqref{1}) to obtain the continuity equation in
terms of the density contrast:
$$
\div\vct{v} = {-\del \over (a-1)t}. \eqno(\new)
$$
The essential point of this paper is made by comparing
equation (\eqref{1}), which is derived assuming that
gravity is non-existent, to equation (5a), which is derived from the
usual linear theory for the growth of perturbations under the influence
of gravity.
Both equations indicate the same relationship between $\div\vct{v}$ and
$\delta$, differing only by the constant of proportionality.  Furthermore,
for $a\sim$ a few (in the non-gravitational case) and $\Omega\sim 1$ (in the
gravitational case), the predicted velocities are of comparable magnitudes.
If the velocity impulse in the non-gravitational model is irrotational,
then the velocity field remains irrotational at later times,
and equation (5b) also holds, apart from the constant of proportionality.

It is equation (5a) or (5b) that is traditionally
used to estimate $\Omega$, given the large-scale velocity and density fields.
In the present case, examination of equation (\eqref{1}) shows that an
observer (or theorist) erroneously applying (5a) or (5b) in an
empty universe would infer an effective value of $\Omega$ determined by
$$ f(\Omega_{eff}) = \left( {1 \over a-1} \right).  \eqno(\new)$$
Even though the true value of the
cosmological density parameter is $\Omega=0$ (\ie there is no gravitating
mass present), one always infers a non-vanishing value for $\Omega_{eff}$.
The solid line in Figure 1 shows the evolution of $\Omega_{eff}$; it is
initially infinite, but it declines
and asymptotically approaches zero.
In this model the ratio of the inferred $\Omega$ to the true $\Omega$ is
always infinite.

\topinsert
\capskip
\vskip 0.4truein
\centerline{\bf FIGURE 1}
\vskip 0.4truein
\medskip
{\eightpoint
\capskip
\noindent {\bf Figure 1} ---
The value of $\Omega$ that would be inferred by an observer applying
equation (5a) or (5b) to the velocity and density fields of
models in which non-gravitational forces,
(\eg explosions or radiation pressure) set the baryons into motion
at $a=a_e$.  The solid line [equation (12)]
represents an $\Omega=0$ model, in
which there are no gravitational forces at all.
The dotted line [equation (19)] represents a two-component
model with $\Omega_B=0.1$ and $\Omega_D=0.9$.
For $a/a_e \sim 2-4$, the non-gravitational scenario has $\Omega_{eff}$
in the observed range even if the true $\Omega$ is zero.
}
\medskip
\textskip
\endinsert

To summarize,
the example of an $\Omega=0$ universe demonstrates that a
proportionality
between the density field and the
divergence of the
velocity field is not solely a property of the
gravitational instability hypothesis.  Analyzing the density
and the velocity fields with equations derived within the framework of this
hypothesis will yield an estimate of $\Omega$, but
unless the motions are gravitational in origin the inferred
value of $\Omega$ is not a measure of the cosmological density parameter.

\bigskip
\centerline{\bf 2.2 A Two-component Universe}
\medskip

Now let us consider a two-component universe consisting of dark matter and
baryons.
We want to study the evolution of the two fluids under
the combined influence of gravitational and non-gravitational forces.
We characterize the mean densities of the dark matter and the baryons
by the parameters $\Omega_D$ and $\Omega_B$, respectively.  For algebraic
simplicity, we shall assume that the total mass density is such that
$\Omega_D + \Omega_B=1$.
We again assume that non-gravitational forces produce an
impulsive perturbation in the velocity field of the baryonic component
at time $t=t_e$,
and that they are negligible at later times.
In the linear regime, the equations
governing the evolution of the baryons and the dark matter
are:\hfil\break
\line{the continuity equations,\hfill}
$$\dt\del_i + \div\vct{v}_i =0, \ \ \ \ i=D,B, \eqno(\new)$$
\line{the Euler equations,\hfill}
$$\eqalign{&\dt\vct{v}_D + 2H(t)\vct{v}_D = - \grad \Phi_g\posx, \cr
           &\dt\vct{v}_B + 2H(t)\vct{v}_B = -\grad \Phi_g\posx
+\vct{V}_B(\vct x) \delta^\dirac(t-t_e), \cr} \eqno(\new)$$
\line{and the Poisson equation,\hfill}
$$\gradsq\Phi_g = 4\pi G\sum_{i=D,B} \bar{\rho}_i(t)\del_i\posx
= {3\over 2} H^2(t)\sum_{i=D,B} \Omega_i\del_i\posx.  \eqno(\new)$$
The comoving gravitational potential $\Phi_g$
mediates the gravitational interaction between the two fluids.
In these equations, the subscripts ``D'' and ``B'' denote dark matter and
baryons, respectively.

By combining the above equations and absorbing the effects of the impulse
perturbation into the initial conditions, we find
that the density perturbations, at time $t > t_e$, evolve according to
$$ \dt^2\del_i +2H(t)\dt\del_i= {3\over 2} H^2(t) \sum_{i^\prime = D,B}
\Omega_{i^\prime}\del_{i^\prime}. \eqno(\new)$$
The general solution to this pair of
differential equations can be expressed as (Hogan \& Kaiser 1989)
$$\left[\matrix{\del_B\cr t\dt\del_B \cr \del_D\cr t\dt\del_D\cr}\right]
= \left[\matrix{1  &\hfill 1  &  1  &  1  \cr
	      2/3&\hfill-1  &-1/3 &  0  \cr
	      1  &\hfill 1  &-\Omega_B & -\Omega_B   \cr
	      2/3&\hfill -1  &-\Omega_B/3 &  0  \cr }\right]
\left[\matrix{\Delta_1(\vct x) a(t)\hfill \cr \Delta_2(\vct x) a(t)^{-3/2}
\hfill\cr \Delta_3(\vct x) a(t)^{-1/2}\hfill \cr\Delta_4(\vct x)\hfill \cr}
\right].\eqno(\new)$$
The spatial functions $\Delta_i(\vct x)$ can be determined by imposing the
appropriate initial conditions and inverting the equation.  Assuming
$\Omega_B \ll 1$, it is
simpler to expand the functions $\Delta_i$ as a series:
$\Delta_i=\Delta_{i0}+\Omega_B\Delta_{i1}+\Omega_B^2\Delta_{i2} +\cdots \,$
and iteratively solve for $\Delta_{i}$ to the desired order.

Our scenario is
constructed along the lines of theories that invoke non-gravitational
forces (such as cosmological explosions
or radiation pressure instabilities)
to initiate the formation of large-scale structure.
We assume that the baryon and dark matter
distributions are smooth at early times, and that the non-gravitational
forces disturb the baryonic component impulsively at $t=t_e$.
The appropriate initial conditions for this scenario
are: $\delta_{D,B}=0$, $\dt\delta_D=0$, and $\dt\delta_B=\zeta
(\vct x)$, where $\zeta(\vct x)\equiv -\div\vct{V}_B(\vct x)$.
At any subsequent time, the baryonic density perturbations are given,
to first order in $\Omega_B$, by:
$$\eqalign{ \delta_B\posx=& 3\zeta(\vct x)t_e\left[ 1 -a^{-1/2} + \frac1/5
\Omega_B\left(a-a^{-3/2}+5a^{-1/2}-5\right)\right], \cr
t\dt\delta_B\posx=&\phantom{3}\zeta(\vct x)t_e\left[ a^{-1/2} + \frac1/5
\Omega_B\left(2a+3a^{-3/2}-5a^{-1/2}\right)\right]. \cr}\eqno(\new)$$
In the limit of $\Omega_B \rightarrow 0$, the initial velocities simply
decay as $a^{-2}$ (in comoving units), and the density contrast
$\delta_B$ asymptotes to $3\zeta(\vct x)t_e$.  When $\Omega_B$ is not
zero, the dark matter eventually
begins to fall into the baryon perturbations, and
the perturbed baryons and dark matter produce gravitational accelerations
parallel to the initial velocities.
These accelerations sustain the velocities and the
growth of the density fluctuations.

For consideration of density-velocity tests,
the crucial feature of equation (\eqref{1})
is that $\dt\del_B(\vct x)$ is proportional to $\del_B(\vct x)$ at
all times.  We can therefore substitute $\del_B$ for $\dt\del_B$ in
the continuity equation (\eqref{6}) and find that
$\div\vct{v}_B \propto -\del_B$, just as in the gravitational instability
scenario.  At all times $\div\vct{v}_B \propto \div\vct{V}_B$, and if
$\vct{v}_B$ is irrotational then we can recover the full velocity field
from its divergence.  The results for this scenario are thus identical to
those for gravitational instability (equations 5a and 5b),
except that the quantity $f(\Omega)$ is replaced by
$$f(\Omega_{eff}) = {1\over 2} \left[ {
a^{-1/2} + \frac1/5\Omega_B\left(2a+3a^{-3/2}-5a^{-1/2}\right)\over
1 -a^{-1/2} + \frac1/5\Omega_B\left(a-a^{-3/2}+5a^{-1/2}-5\right)} \right].
\eqno(\new)$$
The dotted curve in Figure 1 shows the evolution of
$\Omega_{eff}(a)$ for a case where $\Omega_B=0.1$.
The value of $\Omega_{eff}$ is initially very large, but it
drops to its minimum of $0.24$ by $a\approx 10$,
then asymptotically rises to unity.
The initial drop in $\Omega_{eff}$
reflects the decaying influence of the impulse perturbations,
and the return to $\Omega_{eff} \approx 1$ occurs as the dark matter
catches up to the perturbed baryons.
We draw attention to the fact that, with
the exception of a relatively brief period following the impulse,
the value of $\Omega_{eff}$ ranges between $0.24$ and $1$.  Hence,
measurements yielding $\Omega_{eff} < 1$ do not necessarily imply an open
universe.  In the same vein, measurements
of $\Omega_{eff}$ that yield values of ${\cal O}(1)$
{\it do not} preclude the possibility of a non-gravitational influence
having been the original cause of structures in the universe.

Linear theory solutions are more complicated in the case of a universe
with $0<\Omega<1$, but on physical grounds we expect the qualitative results
to be similar.  In particular, the same correlations between density and
velocity fields should appear, just with different constants of
proportionality.
In the non-gravitational scenario, $\Omega_{eff}$ will asymptotically
approach the true value of $\Omega$, though in a two-component
model with small $\Omega_B$ it is possible that the dark matter will
not catch up
by the time the universe enters free expansion,
and that $\Omega_{eff}$ will get stuck between $\Omega_B$ and
$\Omega$.

\bigskip
\centerline{\bf 2.3 Non-linear Effects}
\medskip

The relations in \S 2.1 and \S 2.2
are derived from linearized equations, so they are valid
only in the linear regime.  We can extend the relation between the velocity
and the density fields into the non-linear regime by defining a comoving
displacement field:
$$
\vct{\xi}(\vct{x},t)\equiv\vct{x}(\vct{q},t)-\vct{q}=
t\,\langle{\vct{v}}(\vct{x},t)\rangle_t.
\eqno(\new)$$
Here $\vct{x}(\vct{q},t)$ is the comoving position at time $t$ of a
mass element originally at comoving position $\vct{q}$, and
$\langle{\vct{v}}(\vct{x},t)\rangle_t$ is the time-averaged,
comoving peculiar velocity
of this mass element.  Mass conservation requires that
$\rho_x(\vct{x}) d\vct{x} = \rho_q d\vct{q}$, so
$$
\rho(\vct{x},t)/\bar\rho(t) =
\left\Vert I- {\partial\vct{\xi}\over\partial\vct{x}}\right\Vert ,
\eqno(\new)$$
where $I$ is the identity matrix and vertical bars denote the Jacobian
determinant
(Nusser \etal\ 1991).
Equation (\eqref{1}) remains valid until orbit crossing occurs.  After
this time $\vct{\xi}(\vct{x},t)$ is no longer single-valued, but we
might still expect equation (\eqref{1}) to apply on large scales, after
smoothing over regions of orbit crossing.
In a two-component universe, equation (\eqref{1}) holds separately for
the density and displacement fields of the two components.

If the present-day peculiar velocities are proportional to the time-averaged
peculiar velocities, that is
$\vct{v}(\vct{x},t_0)=\alpha\langle{\vct{v}}(\vct{x},t_0)\rangle_t$, then
equation (\eqref{1}) determines the relation between the present-day
velocity and density fields:
$$
\rho(\vct{x},t_0)/\bar\rho(t_0) =
\left\Vert I- {t_0 \over \alpha}
{\partial\vct{v}\over\partial\vct{x}}\right\Vert .
\eqno(\new)$$
On scales where orbit crossing is unimportant, this equation applies equally
well to gravitational and non-gravitational models, provided only that
the initial matter distribution is uniform and that present and time-averaged
velocities are indeed proportional.  Confirming relation (\eqref{1})
observationally can therefore provide evidence only for this combination
of assumptions, not for the more specific assumption that the displacements
and velocities were generated by gravitational instability.
Furthermore,
comparison of the observed velocity and density fields only yields the
constant of
proportionality relating the present and time-averaged velocities,
which in general need not be related to $\Omega$.

For the case of gravitational instability, Nusser \etal\ (1991) propose
using the Zel'dovich (1970) quasi-linear approximation for the displacements
and the velocities:
$$
\vct{\xi}(\vct{x},t)=D(t)\vct{\psi}(\vct{q}) \qquad \Longrightarrow \qquad
\vct{v}(\vct{x},t)\equiv \dot\vct{\xi} = \dot D(t)\vct{\psi}(\vct{q}),
\eqno(\new)
$$
where $D(t)$ is the growth rate of linear perturbations and
$\dot D/D = H f(\Omega).$  Inserting this approximation into
the mass conservation
equation (\eqref{3}) yields
$$
\rho(\vct{x},t_0)/\bar\rho(t_0) =
\left\Vert I- {D \over \dot D}
{\partial\vct{v}\over\partial\vct{x}}\right\Vert .
\eqno(\new)
$$
In the limit of small $D$, this yields the linear relation
$\delta = -(Hf)^{-1} \div\vct{v}$.
The approximation (\eqref{1}) does not satisfy the Euler and Poisson
equations of gravitational instability theory. Instead, it is a
solution of
the continuity equation under the Zel'dovich {\it ansatz} that the
velocity is proportional to the displacement. Still, Nusser \etal\ (1991)
find, based on $N$-body simulations, that it provides an excellent
approximation to the true velocity-density relation in a
mildly non-linear gravitating system, \ie in the range
$-0.7 \simlt \delta \simlt 4.5$.

One can also devise quasi-linear approximations for computing the
smoothed peculiar velocity field given observed density contrasts
(\eg Nusser \etal\ 1991).  However, non-linear corrections have a
smaller impact when computing velocities from densities,
and we will not consider them in this paper.

The basic message of this section is that
even mildly
non-linear effects do not distinguish
gravitational models from non-gravitational models,
because the most successful approximation in this regime is derived from
the continuity equation and the
proportionality of velocity and displacement, independently of gravity.
Dynamical non-linearities in gravitational systems eventually break the
proportionality between velocities and displacements.
Such deviations from the Zel'dovich approximation
could, in principle, be used
to differentiate gravitational and non-gravitational
theories.  However, the observational data on large-scale velocity flows
are unlikely to permit detection of such subtle effects in the foreseeable
future, and they would in any case be difficult to distinguish from
non-linearities in the relation between galaxy and mass density.
In the highly non-linear regime,
the existence of ``fingers-of-god'' indicates that
gravity plays a major role in the dynamics of galaxy clusters,
but it does not tell us that clusters
formed by gravitational processes alone.

\bigskip
\centerline{\bf 2.4 Biased Galaxy Formation}
\medskip

Observations do not probe the baryon distribution directly; rather,
they map out the galaxy distribution.
It is possible that the galaxy distribution evenly traces the underlying
distribution of baryonic matter, but it is also possible that galaxies
form more efficiently in some regions and less efficiently in others.
The observed morphology-density relation demonstrates that such efficiency
variations occur at least for individual galaxy types.
The idea of ``biased'' galaxy formation --- preferential formation of
galaxies in high density regions --- became popular when it was realized
that such a scheme
could reconcile the assumption of an $\Omega=1$ universe with
observations suggesting lower $\Omega$ (see review by Dekel \& Rees 1987).
Numerical simulations of the cold dark matter model indicate that some
degree of biasing is a natural prediction of this theory
(White \etal~1987; Gelb 1992; Cen \& Ostriker 1992; Katz, Hernquist \&
Weinberg 1992) and, indeed, that physical processes are likely to produce
bias in a wide range of theoretical scenarios.

The simplest mathematical prescription that describes biasing
is the ``linear bias''
model, in which one assumes that the galaxy and baryon density contrasts
are proportional to each other and that the galaxy and baryon velocity
fields are the same, \ie
$\del_G=b_B\del_B$, where $b_B$ is the baryonic ``bias parameter'',
and $\vct v_G= \vct v_B$.
It is clear that our earlier results for the
relations between linear density and velocity fields continue to hold
with this biasing model, except that $f(\Omega_{eff})$ is replaced by
the parameter $\beta\equiv f(\Omega_{eff})/b_B$.  For the gravitational
instability scenario, equations (5) become
$$\eqalignno{
\div\vct{v}_G &= -H\beta\, \del_G\posx, &(\new{\hbox{a}}) \cr
\vct{v}_G\posx &={(H\beta)\over 4\pi} \int d\vct y\ \del_G\posy
{(\vct y-\vct x) \over\left\vert \vct y -\vct x\right\vert^3}.
&(\eqref{1}{\hbox{b}})
}
$$
Similar relations will hold for the sort of non-gravitational model
discussed in \S 2.2, except that the constant of proportionality will
no longer reflect the true cosmological value of $\beta$.

It makes no sense to apply the linear bias model once baryon fluctuations
grow to amplitudes $\del_B \sim 1/b_B$, since it can yield the absurd
prediction that $\del_G<-1$.  We must therefore adopt a different strategy
to incorporate biasing into the quasi-linear relation discussed in \S 2.3.
A simple approach is to put the bias factor inside the determinant
of equation (\eqref{2}), yielding
$$
\del_G(\vct{x})=
\left\Vert I- (H\beta)^{-1}
{\partial\vct{v}_G\over\partial\vct{x}}\right\Vert -1.
\eqno(\new)
$$
In the linear limit, this yields equation (\eqref{2}a).
Some of our numerical simulations of gravitational models (see \S 3.1)
employ a biasing scheme in which galaxies are identified with high peaks
of the initial density field, and for these simulations we find that
putting the bias factor inside the determinant yields better correlations
than a simple linear bias model.  There is no guarantee, however, that this
prescription will work for other biasing schemes.

The cosmic mass distribution must obey the continuity equation, but the
galaxy distribution need not.  In a biased model, structure in the galaxy
distribution is not created by displacing an initially uniform population
of objects; it also reflects variations in the efficiency of galaxy
formation from one place to another.
The continuity argument presented at the beginning of this section does
not apply directly to the galaxy distribution in such a model.
Nonetheless, to the extent that the relation between galaxies and mass
is described by linear bias, the standard relation between velocity and
density will continue to hold, with the $1/b_B$ change in the constant
of proportionality.  Mild non-linearities in the bias relation may
be difficult to distinguish from kinematic or dynamical non-linearity
in the velocity-density relation.  However, if the connection between galaxies
and mass is non-local, then the resulting model can fail density-velocity
tests in a drastic way, even if the mass distribution and the velocity field
grow entirely by gravitational instability.  We return to this issue in \S 5.

\bigskip
\centerline{\bf 3. Simulations}
\bigskip

The arguments in the previous section are somewhat idealized, as they
are based largely on linear theory and the linear biasing model.
In the following three sections, we illustrate these arguments with
concrete examples derived from $N$-body simulations of gravitational
and non-gravitational models.
Although the \denvel and \velden techniques discussed in \S 1
have been tested on $N$-body simulations of gravitational models in other
papers (Nusser \etal~1991; Davis \etal~1991; Dekel \etal\ 1993),  we repeat
the exercise here because we
want to compare our results for explosion simulations to results for
gravitational simulations analyzed with identical techniques on the same
scales. We intend to demonstrate
that the correlations between velocity and density fields discussed in
\S 2 can be found in realistically non-linear models of the universe,
regardless
of the mechanism that caused the velocity flow.
We also want to investigate the impact of biased galaxy
formation, which has not been well studied in previous work.

\bigskip
\centerline{\bf 3.1 Gravitational Models}
\medskip

For our gravitational simulations, we assume an $\Omega=1$ universe and
adopt the cold dark matter (CDM) power spectrum (Bardeen \etal~1986) for the
initial fluctuations in the mass distribution.
The adopted CDM power spectrum is an adequate representation of reality
for the purposes of this paper, even though, in detail, it may fail
rigorous tests of its applicability to the observed universe
(\cf Ostriker 1993 and references therein).
We use the Zel'dovich
approximation (Zel'dovich 1970)
to transform the initial conditions into positions and growing mode
velocities of the particles, and evolve the resulting distribution
using a particle-mesh (PM) $N$-body code
written by C.\ Park (Park 1990).  This code uses a staggered-mesh
technique (Melott 1986) to achieve higher force resolution (by about a
factor of two) than a conventional PM code.
Tests against analytical solutions and other $N$-body codes
show that the PM code produces reliable results down to the limits of
its force resolution, $\sim 1-2$ mesh cells
(Park 1990; Weinberg \etal, in preparation).
The simulations of gravitational models
employ $64^3$ particles on a $128^3$
mesh representing a periodic cube of length $96h^{-1}\mpc$,
for a Hubble constant of $H_0=50\kms\mpc^{-1}$.
Table 1 lists more details of the simulations.

\topinsert
\centerline{Table 1a: Parameters of Gravitational Simulations}
\bigskip
\hrule\vskip0.1truecm\hrule
$$\vbox{\tabskip 1em plus 2em minus 5em
\halign to\hsize{ # \hfil & # \hfil & \hfil # & \hfil # \hfil &
\hfil # \hfil & \hfil # \hfil & \hfil # \hfil& \hfil # \hfil \cr
{\hfil Model\hfil} &&$L$&$a_f/a_i$&$N_t$&$N_p$&$N_m$&
\cr
\noalign{\medskip\hrule\medskip}
\undertext{CDM} & Unbiased & 96 & 24  &  48  & $64^3$  & $128^3$\cr
                & Biased &   96 & 12  &  24  & $64^3$  & $128^3$\cr
\noalign{\medskip\hrule}
}}$$
\smallskip
\vskip -12pt
{\eightpoint
\capskip
\line{Notes to Table 1a: \hfil}
\line{Column 3: size of simulation cube, in $h^{-1}$ Mpc \hfil}
\line{Column 4: ratio of final to initial expansion factor \hfil}
\line{Column 5: number of timesteps (equal intervals in $\Delta a$) \hfil}
\line{Column 6: number of particles \hfil}
\line{Column 7: number of cells in density/potential mesh \hfil}
}
\bigskip
\capskip
\vskip 0.4truein
\centerline{\bf FIGURE 2}
\vskip 0.4truein
\medskip
{\eightpoint
\capskip
\noindent {\bf Figure 2} ---
Slices from our $N$-body simulations of the cold dark matter (CDM) model
(see Table 1a for defining parameters).
Each column shows three contiguous slices from a single simulation.
Each slice is $1/6$ the thickness of the simulation cube, so that one half
of the simulation volume is shown in total.  The scale is marked
in $\hmpc$.
The left-hand column shows slices
from an unbiased simulation.  The middle column
shows slices through the galaxy distribution of a biased simulation.
The right-hand column shows slices through the mass distribution of the
same biased simulation.  This mass distribution is sampled to the same mean
density as the galaxy distribution, so that visual differences are not
dominated by differences in particle density.
}
\textskip
\endinsert

Since the $N$-body simulations only
follow the evolution of the mass, the identification of
galaxies in the simulations requires
further assumptions.
The usual schemes for identifying galaxies
either assume that galaxies trace the matter (the unbiased scenario,
corresponding to bias parameter $b=1$) or implement an {\it ad hoc}
prescription to model biased galaxy formation, such as flagging
high peaks in the initial density field as sites for galaxy formation
(Bardeen \etal~1986).  We explore both schemes.
For our unbiased models we adopt the standard normalization condition
that the rms fluctuation of the initial density field linearly
extrapolated to $z=0$
be
unity in spheres of radius $8 h^{-1}\mpc.$
We create a galaxy catalog by randomly sampling the particle distribution
to a density of $0.01h^{3}\mpc^{-3}$,
the observed number density of bright ($L_*$-ish) galaxies.
We normalize our biased models so that the rms linear mass
fluctuation is 0.5 at $8 h^{-1}\mpc$, and we use the peak-background
split scheme (Bardeen \etal~1986; Park 1991) to identify biased particles
corresponding to $\nu > 1.6\sigma$ peaks of the initial conditions
on a scale of $0.6 h^{-1}\mpc.$
With this choice of peak parameters, the mean galaxy density is again
$0.01h^{3}\mpc^{-3}$, and the bias factor is $b=2$.  The rms
galaxy fluctuation in $8 h^{-1}\mpc$ spheres is, therefore, unity, as in
the unbiased model.
Strictly speaking, the galaxy distribution in the peak-biased scenario
does not satisfy the continuity equation because the galaxies are ``born''
clustered.  However, the fluctuations in the final galaxy density field are
closely coupled to the fluctuations in the mass density,
which does obey the continuity equation.

Figure 2 shows slices through the galaxy distributions of the
gravitational models.
The three panels in each column display three contiguous
slices from a single simulation.  Each slice is $1/6$ the thickness of
the simulation cube, so that one half of the simulation is shown in total.
The left-hand column shows slices from one of
the unbiased simulations.  The middle column shows slices
from the biased simulation with the same initial fluctuations.
The galaxy distributions of the two models look similar, although structure
in the unbiased model is clumpier, while the biased model
has slightly larger voids and a somewhat more filamentary appearance.
Of course, peculiar velocities are lower in the biased model
(by a factor of two in the linear regime) because the amplitude of
mass fluctuations is lower.  The right-hand column of Figure 2 shows
the mass distribution of the biased model.  We have sampled the
mass distribution to $0.01 h^3\mpc^{-3}$ so that the visual differences
between the mass and galaxy distributions are not dominated by
differences in the particle densities.

For the purposes of this paper, it is not terribly important whether the
peaks biasing scheme gives an accurate description of galaxy formation
in the CDM model, or whether the CDM models or the explosion models reproduce
all features of the observed galaxy distribution.
We are interested in broader questions about scales that are mildly
non-linear, and it is sufficient that our simulations yield qualitatively
realistic structure while illustrating the basic effects of
gravitational instability, biased galaxy formation, and non-gravitational
perturbations.

\bigskip
\centerline{\bf 3.2 Explosion Models}
\medskip

For our non-gravitational models, we use numerical simulations of the
explosion scenario.  These simulations are akin to those described by
Saarinen, Dekel \& Carr (1987), but we use different numerical
techniques, many more particles, larger physical volumes, and a wider
range of initial conditions.  A detailed discussion of the simulations
and statistical analysis of the structure that they produce will
appear elsewhere
(Weinberg, Dekel \& Ostriker, in preparation);
here we provide a summary of the models and the numerical methods.
The explosion scenario serves here as a representative example of a
non-gravitational theory of structure formation --- a convenient
example because we happen to have simulations of it.
However, our goal is to learn about the relation between velocity and
density fields, not to defend or refute the explosion model, so we will
not attempt any comprehensive evaluation of the model's observational
successes or failures.

The detailed physics of the explosions is not particularly important
for our purposes.  We simply assume that some explosive process sweeps
baryonic material onto expanding, spherical shells.
We assume that, in the absence of interactions, the shells would have a
power-law distribution of radii up to some maximum,
$n(R) \propto R^{-4}$ for $R \leq R_{max}$.
Ostriker \& Strassler (1989) and Weinberg, Ostriker \& Dekel (1989)
have argued that a model with an $R^{-4}$ distribution can produce
a reasonable match to the distribution of void sizes in the CfA slice
and to the distribution of galaxy cluster masses, respectively.
The physical scale corresponding to $R_{max}$ is treated as a
free parameter, to be determined {\it a posteriori} by fitting the
observed level of galaxy clustering.
We have also examined an explosion model in which all shells have equal radii;
results are similar to those reported here for the power-law model.

We consider three different cosmological scenarios.  In the first, we
set $\Omega_S=\Omega=1$, where $\Omega$ is the density parameter and
$\Omega_S$ represents the cosmological density of the material that
can be swept up onto shells.  One would normally expect blast waves
to collect only baryonic matter, and setting $\Omega_B=\Omega_S=1$
would violate nucleosynthesis constraints.  However, $\Omega_S=1$ is
possible if explosions occur early, so that dark matter has time to
catch up, or if the ``explosions'' have a gravitational origin, with
expanding shells developing around deep negative density perturbations
(\cf Bertschinger 1985; Weinberg \& Cole 1992).
Our second scenario assumes an open, baryon-dominated universe,
with $\Omega_S=\Omega=0.17$ at redshift $z=0.$  Finally, we consider
a scenario with $\Omega=1$ and $\Omega_S=0.1$, \ie 90\% of the mass
is assumed to be in a collisionless dark matter component that is not
directly disturbed by the explosions, though it can respond gravitationally
to the perturbed baryon distribution.
For brevity, we will refer to these
three scenarios as ``flat'' ($\Omega_S=\Omega=1$),
``open'' ($\Omega_S=\Omega=0.17$), and ``dark'' ($\Omega_S=0.1, \Omega=1$),
respectively.

We evolve the explosion models in two phases, a ``hydrodynamic'' phase
and a ``gravitational'' phase.  Our idealization is that gravitational
effects can be ignored during the early evolution of the blast waves,
the hydrodynamic phase, but that after a certain point (physically
speaking, the point at which the shells begin to fragment), the
evolution enters a gravitational phase during which matter is
effectively collisionless and hydrodynamics can be ignored.
Because the large-scale
growth and interactions of cosmological shells are similar regardless of
whether they are collisional or collisionless, our results should not be
sensitive to this idealization.
Of course the fragmentation process must involve gravity, but it
does so on scales much smaller than those that we are considering, where
the density field is strongly non-linear.

The hydrodynamic phase would be simple to compute (for our rather simplified
purposes) if shells did not overlap: we would simply choose random locations
for the shell centers (the ``explosions''), project particles lying within
the shell radius out to the surface, and assign them radial peculiar
velocities.  However, overlap regions are unavoidable,
and they are tricky to manage ``by hand.''
Therefore, we turn to a technique based on
Burgers' equation (Burgers 1974), which describes the flow of a
fluid with bulk viscosity.  Burgers' equation was introduced into cosmology by
Gurbatov, Saichev \& Shandarin (1985, 1989), who used it as a clever extension
of the Zel'dovich approximation for gravitational instability.
Weinberg \& Gunn (1990) describe an efficient technique for integrating
the equation in three dimensions.
We employ this technique here,
but instead of regarding Burgers' equation as an approximation for
gravitational instability, we return to its original, hydrodynamic
interpretation and treat it as a solution for the purely
hydrodynamical interactions of a system of blast waves.
Given an initial velocity field that is the gradient of a potential,
we can compute the velocity field at any later time and integrate
particle orbits along this evolving velocity field.
A spike in the initial velocity potential
drives a spherical outflow,
which, because of the viscosity term in Burgers' equation, sweeps all
particles in its path onto a thin shell, whose radius grows with time.
Where two shells emanating
from neighboring spikes collide, viscosity prevents
the shells from interpenetrating, and it redirects particle velocities
in a momentum-conserving way along the wall that separates the
two bubbles.  Burgers' equation thus
automatically handles blast wave interactions in a physically plausible manner.

We place the potential spikes (``explosion'' sites) at random locations
in a periodic, cubical simulation volume $V$,
which contains $64^3$ particles representing the baryonic component
(or, more generally,
the ``shell'' component).
The height of a potential spike
determines the radius of the associated shell, in cube units.
We distribute the spike heights so that, in the absence of shell collisions,
the distribution of shell sizes would be $n(R) = n_0 R^{-4}$ for
$R_{max} \geq R \geq R_{max}/16.$  By the end of the hydrodynamic phase,
many of the smaller bubbles have been swept up or crushed between larger
shells.  The constant $n_0$ is chosen so that
the formal filling factor is
$$
ff \equiv
{1 \over V}\int_{R_{max}/16}^{R_{max}} {4\pi \over 3}R^3 n(R)\,dR ~=~ 3.
\eqno(\new)
$$
The formal filling factor of large shells
($R_{max} \geq R \geq R_{max}/2$) is 0.75.
The choice of filling factor defines the end of the hydrodynamic phase ---
shells grow hydrodynamically until they fill the specified volume.
Before beginning the gravitational phase, we multiply the Burgers' equation
velocity field by a constant factor, chosen so that the peculiar expansion
velocity of an isolated, undisturbed shell would be 20\% of its Hubble
expansion velocity, $\vct V_p=0.2H\vct R$.  In principle, the expansion
velocity is another parameter of the model; we choose the value implied by
the self-similar solution in an $\Omega_S=\Omega=1$ scenario (Bertschinger
1985), which also seems reasonable for our other cases.

\topinsert
\centerline{Table 1b: Parameters of Explosion Simulations}
\bigskip
\hrule\vskip0.1truecm\hrule
$$\vbox{\tabskip 1em plus 2em minus 5em
\halign to\hsize{ #\hfil & \hfil # & \hfil # &
\hfil # & \hfil # & \hfil # & \hfil # & \hfil # & \hfil #  \cr
{\hfil Model\hfil} &$\Omega$&$\Omega_S$&$L$&$ff(a_h)$&$a_f/a_h$&
$N_t$&$N_p$&$N_m$\cr
\noalign{\medskip\hrule\medskip}
Flat
& 1.0 & 1.0 & 108 & 3.0  & 2 & 16 & $64^3$ & $128^3$\cr
Open
& 0.17 & 0.17 &  98 & 3.0  & 4 & 32 & $64^3$ & $128^3$\cr
Dark
& 1.0 & 0.1 &  98 & 3.0  & 6 & 48 & $2 \times 64^3$ & $128^3$\cr
\noalign{\medskip\hrule\medskip}
}}$$
\smallskip
\vskip -12pt
{\eightpoint
\capskip
\line{Notes to Table 1b: \hfil}
\line{Column 2: density parameter \hfil}
\line{Column 3: density parameter of ``shell'' component \hfil}
\line{Column 4: size of simulation cube, in $h^{-1}$ Mpc \hfil}
\line{Column 5: formal filling factor at end of hydrodynamic phase \hfil}
\line{Column 6: ratio of final expansion factor to expansion factor at end
of hydrodynamic phase\hfil}
\line{Column 7: number of timesteps (equal intervals in $\Delta a$) \hfil}
\line{Column 8: number of particles; in dark simulations there are $64^3$
particles of each species\hfil}
\line{Column 9: number of cells in density/potential mesh \hfil}
}
\medskip
\textskip
\endinsert

For the gravitational phase of the simulations, we take the particle positions
and velocities at the end of the hydrodynamic phase as initial conditions
and evolve them forward using Park's PM code.
We again use $64^3$ particles on a $128^3$ mesh.
For the dark models we use an additional $64^3$ particles
to represent the dark component.  These particles are uniformly distributed
and unperturbed at the beginning of the gravitational phase, but they
quickly start to chase after the baryon perturbations.

\topinsert
\capskip
\vskip 0.4truein
\centerline{\bf FIGURE 3}
\vskip 0.4truein
\medskip
{\eightpoint
\capskip
\noindent {\bf Figure 3} ---
Slices from our $N$-body simulations of the explosion scenario
(see Table 1b for defining parameters).
Each slice is $1/6$ the thickness of the simulation cube, and each
column displays three contiguous slices.
The first three columns show, from left to right, the
galaxy distributions of the flat model ($\Omega_S=\Omega=1$), the
open model ($\Omega_S=\Omega=0.17$),
and the dark model ($\Omega_S=0.1,\,\Omega=1$).
The fourth column shows the distribution of dark mass in the dark model.
In the absence of collisions, the shells would have a
power-law radius distribution, $n(R) \propto R^{-4}$.
}
\medskip
\textskip
\endinsert

At the end of the hydrodynamic phase, the two-point correlation functions
of the particle distributions are roughly $r^{-1}$ power-laws on small
scales, the $-1$ power-law index reflecting the two-dimensional nature of the
dominant structures, the shells.  Gravitational clustering causes the
correlation functions to steepen on small scales.  We identify the
present epoch as the time when the correlation function most closely
matches the observed $r^{-1.8}$ power-law.  This occurs after an expansion
factor of two (following the start of the gravitational phase) for the flat
model, four for the open model, and six for the dark model.
We fix the size of the simulation box (in $h^{-1}$ Mpc)
by requiring that the rms
fluctuation of the baryon particle distribution in spheres of radius
$8 h^{-1}$ Mpc match the observed galaxy clustering amplitude $\sigma_8
\approx 1$ (Davis \& Peebles 1983).
The simulation boxes turn
out to be $108 h^{-1}$ Mpc for the flat model and $\sim 98 h^{-1}$
Mpc for the open and dark models.  If the shells in the simulations
had evolved without interacting with their neighbors,
the largest shells would have reached
$R_{max}\sim 35-40\, h^{-1}\mpc$.  However,
the volume filling factor of the explosions is sufficiently high that all
the shells are likely to have their growth slowed by collisions
with other shells.  Finally, to create ``galaxy'' distributions,
we can adopt either of two approaches.  We can assume that galaxies evenly
trace the baryonic mass, in which case we just randomly
sample the baryon particle distributions to a density of $0.01 h^3$ Mpc$^{-3}$,
or we can assume that the high gas density of a blast wave shock is
a necessary prerequisite for galaxy formation (see Ostriker \& Cowie 1981;
Vishniac 1983), in which case we randomly sample only those baryon
particles that have been swept onto shells before the end of the
hydrodynamic phase.  With our assumed power-law distribution, the explosions
completely fill space (except for a tiny fraction left undisturbed because
of our arbitrary lower cutoff $R_{min}$), so all baryons are ``on-shell'',
and the two approaches are equivalent.  If the shells do not fill space,
then the two assumptions yield different results, as we discuss in \S 5.
Additional information about the explosion $N$-body
simulations appears in Table 1b.

An interesting feature of Burgers' equation is that, if the velocity
field is irrotational at the initial time, then it remains irrotational at
all later times.  At the end of the hydrodynamic phase, therefore,
the velocity fields of our simulations are irrotational.
Since gravity does not generate large-scale vorticity,
we can guess that the model velocity fields
will remain approximately irrotational on large scales even during
the gravitational phase, and we find this to be the case in our simulations.
To the extent that Burgers' equation gives a reasonable description of
the interactions of hydrodynamic blast waves, we expect this
low vorticity
to be a generic feature of the explosion scenario.

Figure 3 shows slices through the galaxy distributions of the explosion
models, in a format similar to Figure 2.
The first column displays the flat model, the second column the open
model, and the third column the dark model.
The three galaxy distributions look quite similar because we choose the
same sites for the ``explosions'' in each case and because our
normalization procedure ensures that the small scale clustering as
measured by the two-point correlation function is similar in the
three models.  The fourth column shows the dark matter distribution in
the dark model.  The dark matter has begun to reflect the pattern in the
baryon distribution, but a substantial fraction of it remains in the voids.
The range of shell sizes in these models gives the structure an irregular
appearance, with qualitative variations from slice to slice, in rough
agreement with the observed galaxy distribution.

\bigskip
\centerline{\bf 4. Density-Velocity Analysis}
\bigskip

To analyze the simulations, we must transform the final galaxy positions
and velocities into smooth density and velocity fields.
Smoothing removes gross non-linearities, and in any practical situation
it is also needed to reduce shot noise errors.
Smoothing is part of the POTENT \velden procedures from start to finish.
It is only partly implemented in existing
observational applications of \denvel analysis,
where velocities predicted from the smoothed density
field are usually compared to estimated velocities of individual galaxies,
unsmoothed.  In this paper we will adopt the more physically appropriate
procedure of comparing the predicted velocity field to the ``observed''
velocity field smoothed at the same scale, though we recognize that this
approach is trickier to implement in the real universe.

Creating a smoothed density field
from the galaxy positions
in our periodic simulations is straightforward: we use cloud-in-cell
interpolation to compute a density field on a grid and smooth it by
Fourier convolution with a Gaussian.  Creating a smoothed velocity field
is more problematic because in low density regions there may be no
galaxy particles available to trace the local flow, leaving the velocity
field undefined.  The galaxy momentum field
is well defined everywhere, however; where there are no galaxies
the momentum is zero.  We therefore
{\it define} the smoothed velocity field to be the ratio of the
smoothed momentum field to the smoothed density field.
Contributions to the velocity field are thus weighted by
galaxy density. Other smoothing schemes can be used to achieve
something close to volume-weighted smoothing (\S 6 below, also Nusser
\etal~1991).  In the linear regime,
the various velocity smoothing schemes
should be equivalent if galaxies trace mass.
In general, however,
different schemes
will yield somewhat different results.

We should note that in the non-linear regime, galaxy-weighted
smoothing can introduce artificial vorticity into the smoothed
velocity field because (1) smoothing tends to mix streamlines, and (2)
the galaxies may trace an underlying curl-free field unevenly.
In the models to be examined, however, we find that the
vorticity of the smoothed fields
is small, and, to a good approximation, the
velocity field traced by the galaxies can be assumed to be irrotational.

In this paper we present results for Gaussian smoothing lengths
of 1200 km s$^{-1}$
and 600 km s$^{-1}$.  The larger scale is typical of the smoothing now
used for the POTENT \velden analyses (\eg Dekel \etal\ 1993).
As more and better peculiar velocity
data become available, increasing the sampling density, it may be possible
to reduce the smoothing scales to $\sim 600$ km s$^{-1}$,
at least in selected nearby regions.

\bigskip
\centerline{\bf 4.1 Recovering the Density Field}
\medskip

We begin by comparing the true density fields
with those reconstructed from
the corresponding velocity fields.
We  reconstruct the density fields according to one of the
following prescriptions:
$$\eqalign{\delta_{G1} =& -(H\beta)^{-1}\,\div\vct{v}_G,\cr
	   \delta_{G2} =&	\left\Vert I-(H\beta)^{-1}\,{\partial\vct{v}_G
                                \over \partial\vct{x}}\right\Vert -1.\cr}
					         \eqno(\new)$$
For the linear theory reconstruction $\delta_{G1}$ (\cf equation \eqref{4}a),
we first set the ratio $\beta \equiv f(\Omega)/b$
to unity and take the divergence of the
smoothed velocity field to obtain $\delta_{G1}.$
Dividing the rms fluctuation of the reconstructed density
field by that of the simulation density field then yields the
estimated value of $\beta$.
The reconstruction $\delta_{G2}$ (\cf equation \eqref{3} and the
accompanying discussion) uses the quasi-linear relation derived
from the Zel'dovich approximation by Nusser \etal~(1991).
In computing $\delta_{G2}$, we
determine the value of $\beta$
by iterating until the rms density contrast of
the reconstructed field matches that of the simulation field.

\bigskip
\centerline{\bf 4.1.1 Gravitational Models}
\medskip

Figure 4 juxtaposes
a contour plot of a single slice through the smoothed
galaxy density field of one of the unbiased gravitational simulations
against corresponding slices from the two reconstructed density fields,
derived from the smoothed galaxy velocity field by equation (\eqref{1}).
The smoothing scale is $R_s=1200\kms$.  The reconstructed density fields
have been scaled so that their rms fluctuations match that of the
simulation density field.
The heavy contour traces the mean density, while lighter
solid (dotted) contours map out regions of positive (negative) density
contrasts.  The contour interval is $0.2$ (in $\delta$).
The location of the slice is in the middle of the range covered by
the upper left panel of Figure 2.

A comparison of the contour maps shows that the linear theory
reconstruction $\delta_{G1}$ tends to underestimate the magnitude of the
density contrasts in high density regions.
The quasi-linear reconstruction $\delta_{G2}$
fares somewhat better, as it should.
However, the rms density contrast of the fields is only 0.22, so the
differences between the two reconstructions are not large.

\topinsert
\capskip
\vskip 0.4truein
\centerline{\bf FIGURE 4}
\vskip 0.4truein
\smallskip
{\eightpoint
\capskip
\noindent {\bf Figure 4} ---
Velocity-to-density reconstruction of an unbiased gravitational simulation.
The left panel shows a slice through the smoothed galaxy density field of
an unbiased CDM simulation, with a Gaussian smoothing length
$R_s=1200\kms$.  The heavy solid line marks the mean density contour;
positive (solid) and negative (dotted) contours are separated by 0.2
in $\del$.  The central panel shows a slice through the density field
recovered from the smoothed velocity field using the linear
approximation $\del_{G1}$ of equation (28).  The right-hand panel shows
the density field reconstructed via the quasi-linear approximation $\del_{G2}$.
Both methods recover the true density field quite accurately.
}
\bigskip
\medskip
\vskip 0.4truein
\centerline{\bf FIGURE 5}
\vskip 0.4truein
\medskip
{\eightpoint
\capskip
\noindent {\bf Figure 5} ---
($a$)
Scatterplots of the \velden reconstructions of unbiased gravitational models.
Each panel plots the reconstructed density contrast against the true
density contrast ($\del_G$) for
a set of 4096 pixels uniformly spaced throughout the simulation volume.
Left hand panels show a smoothing length of $1200 \kms$, right-hand panels
a smoothing length of $600 \kms$.  Upper panels show the linear theory
reconstruction $\del_{G1}$, lower panels the quasi-linear reconstruction
$\del_{G2}$ (equation 28).  Fluctuations are normalized by the rms
fluctuation $\sigma$ of the corresponding density field.
Diagonal lines mark the $y=x$ relation expected for a perfect reconstruction.
The points follow this relation with modest scatter and some curvature;
the latter is most noticeable in the linear reconstructions and at the
smaller smoothing length.
Values of the linear correlation coefficient $r$ are listed in each panel.
($b$) Same as ($a$), but for biased gravitational models.
}
\textskip
\endinsert

The scatterplots in
Figure 5a present a more quantitative comparison between the true
and reconstructed density fields of this simulation.
For a set of 4096 pixels uniformly spaced throughout the simulation volume,
we plot the pixel's density contrast in the reconstructed field against
its density contrast in the true smoothed density field.
We normalize the density contrasts by the rms fluctuation of the
corresponding field; these rms values are listed in each panel.
The left-hand panels show results for a smoothing length of $1200 \kms$,
and the right-hand panels show results for $600 \kms$ smoothing.
Upper panels show the linear theory reconstruction $\delta_{G1}$, lower panels
the quasi-linear reconstruction $\delta_{G2}$.
A perfect reconstruction would yield equal rms fluctuations in the true
and reconstructed fields and a set of points lying on the $y=x$ diagonal,
marked in each panel by a solid line.
While all of the plots show scatter and some show curvature,
they also display clear correlations, regardless of the reconstruction scheme
or the smoothing scale.
As a quantitative measure, we list in each panel the linear correlation
coefficient,
$r=\sum_i (x_i-\bar x) (y_i-\bar y) /(\sigma_x\sigma_y)$,
where $x$ and $y$ refer to the quantities plotted on the $x$- and $y$-axes,
respectively.
A value $r=1$ corresponds to a perfect correlation,
$r=0$ to no correlation, and $r=-1$ to perfect anti-correlation
(see Press \etal\ 1992 for further discussion).
The values in Figure 5a range from 0.84 to 0.96.

Focusing first on the left-hand panels, we see that the linear theory
reconstruction (top panel) overestimates the amplitude of deep negative
density contrasts and underestimates the amplitude of high positive
density contrasts, so that the scatterplot exhibits significant curvature.
Furthermore, the rms density contrast of the reconstructed
density field, $\sigma_{G1}$, is lower than that of the true density field,
$\sigma_G.$  The quasi-linear method (lower panel) removes the curvature
in the scatterplot, yielding a fairly tight, linear correlation between
the reconstructed and true density contrasts.  The rms contrast $\sigma_{G2}$
equals $\sigma_G$ by construction, since we iterate the assumed value
of $\beta$ until these two values match.
Results for the $600 \kms$ smoothing length (right-hand panels) are
generally similar, although there is greater scatter in the reconstructed
densities, the curvature in the linear theory scatterplot is more severe,
and even the quasi-linear scheme tends to underestimate the highest
densities in the simulation.
The plots in Figures 4 and 5 are all derived from a single run; results for the
other two gravitational runs are similar.

Table 2a lists the estimates of $\beta$ derived from these reconstructions
(averaged over the three independent simulations),
as well as the true values.
The linear theory estimate $\beta_1$, in column 4, is simply the ratio
$\sigma_{G1}/\sigma_G$.  The quasi-linear estimate $\beta_2$, in column 5, is
determined by iterating the value of $\beta$ in equation (\eqref{1}) for
$\delta_{G2}$ until the rms fluctuation of the reconstructed density field
matches the measured value.
The inferred values of $\beta$ are always smaller than the true value of
$\beta=1$, and the linear theory estimates are lower than the quasi-linear
estimates.  These trends hold separately in each of the three simulations,
as well as in the average results.
At $1200 \kms$ smoothing, linear theory yields $\beta_1=0.87$,
and the quasi-linear estimate yields $\beta_2=0.90$; the latter value
indicates that even the Zel'dovich approximation cannot completely remove
the effects of non-linearity.  Non-linear effects are more severe at
the $600\kms$ smoothing length, and the resulting $\beta$ estimates are lower,
$\beta_1=0.74$ and $\beta_2=0.85$.
Despite the systematic tendency to underestimate $\beta$, all of the inferred
values correspond reasonably well to the true value.

\bigskip
\centerline{Table 2a: Inferred $\beta$ values for Gravitational Models}
\bigskip
\hrule\vskip0.1truecm\hrule
$$\vbox{\tabskip 1em plus 2em minus 5em
\halign to\hsize{ # \hfil & # \hfil & \hfil # & \hfil # \hfil &
\hfil # \hfil & \hfil # \hfil & \hfil # \hfil& \hfil # \hfil \cr
{\hfil Model\hfil}\span &$R_s$&$\beta_{true}$&$\beta_1$&$\beta_2$&$\beta_1$&
\cr
&&&       & {\hfil \velden \hfil}\span & {\hfil \denvel
\hfil} \cr
\noalign{\medskip\hrule\medskip}
\undertext{CDM} & Unbiased & 1200 & 1.0 & 0.87 &  0.90  &  0.92 &  \cr
                        & &  600 &  1.0 & 0.74 &  0.85  &  0.87 &  \cr
               & Biased   & 1200 &  0.5 & 0.46 &  0.47  &  0.48 &  \cr
                        & &  600 &  0.5 & 0.42 &  0.48  &  0.46 &  \cr
\noalign{\medskip\hrule\medskip}
}}$$
\smallskip

Figure 5b shows scatterplots of true and reconstructed densities for a
biased CDM simulation, with the same format as Figure 5a.
Results are similar to those for the unbiased model,
and therefore noteworthy.  As mentioned previously,
the galaxy distribution in the peak-biased scenario does not
satisfy the continuity equation, since the galaxies are ``born''
clustered.
However, the fluctuations in the final galaxy density field
are roughly proportional to the fluctuations in the underlying mass density,
which does obey continuity equation, hence the correlations between
density and velocity.
The rms density contrasts of the linear theory reconstructions,
$\sigma_{G1}$, are lower than the corresponding values of $\sigma_G$ by
roughly a factor of two, since the divergence of the velocity field
yields an estimate of the mass density contrast rather than the (biased)
galaxy density contrast.  The linear theory scatterplots also show
significant curvature, especially at $R_s=600\kms.$
The quasi-linear reconstructions are iterated to yield $\sigma_{G2}=\sigma_G$,
and this procedure removes the curvature in the correlations.
This result is encouraging, since it suggests that the quasi-linear
reconstruction scheme, designed to account for non-linear effects of
gravitational evolution, can also correct for departures from linear biasing.
However, we have tested this procedure only for our adopted prescription
of peaks biasing, and it is not clear how well the result will generalize
to other biasing prescriptions.

The last two rows of Table 2a list the values of $\beta$ estimated from
the biased CDM simulations.  Since $\Omega=1$ and $b=2$, the true value
in the simulations is $\beta\equiv f(\Omega)/b=0.5.$
The derived values are slightly lower than this $(0.42\leq\beta\leq 0.48)$,
and again the non-linear estimates are more accurate.

\bigskip
\centerline{\bf 4.1.2 Explosion Models}
\medskip

Figure 6 illustrates \velden reconstructions of an explosion simulation.
The left-hand contour plot shows a slice through the
smoothed density field of one of the flat ($\Omega_S=\Omega=1$)
model simulations, and
the right-hand plots show slices through the two reconstructed density fields.
The location of the slices corresponds to the middle of the range covered
by the upper left panel of Figure 3.
The smoothing scale is $R_s=1200\kms$, and the contour interval
is 0.2 (in $\delta$).  The reconstructed density fields have been scaled
so that their rms fluctuations match that of the simulation density field.

\topinsert
\capskip
\vskip 0.4truein
\centerline{\bf FIGURE 6}
\vskip 0.4truein
\medskip
{\eightpoint
\capskip
\noindent {\bf Figure 6} ---
Velocity-to-density reconstruction of a
simulation of the flat ($\Omega_S=\Omega=1)$ explosion model,
with a smoothing length of $1200 \kms$.  The format is the same as Figure 4:
true density contrast on the left, linear theory reconstruction in
the middle, quasi-linear reconstruction on the right, contour spacing of 0.2
in $\delta$.  Non-gravitational forces play a major role in this model,
but the reconstruction is still quite successful because the
galaxy distribution obeys the continuity equation.
}
\medskip
\textskip
\endinsert

As in the gravitational reconstructions, linear theory tends to underestimate
the density contrasts in high density regions.
More seriously, however, linear theory reconstruction overestimates the
magnitude of density contrasts in low density regions, producing
excessively deep negative perturbations.
The quasi-linear reconstruction fares better in both the high and
low density regions, though it is still not perfect.
Part of the discrepancy reflects the non-linear nature of the density
contrasts in the voids, but another problem arises from noise in
the smoothed simulation velocity field used to reconstruct the density field.
Because there are very few tracer particles (galaxies) in the voids,
the smoothed velocity field is not well defined in these regions.

Figure 7 presents scatterplots of the density fields contoured in Figure 6.
The format is the same as that used for the gravitational models in Figure 5.
The scatterplots of the explosion model look somewhat different
because the 1-point probability distribution is different:
relative to the gravitational models, the very low density regions in the
explosion model occupy a larger fraction of the volume, while the very
high density regions occupy a smaller fraction.
In spite of these differences, and in spite of the fact that the structure in
this simulation has been sculpted primarily by the explosions,
there is a good overall correlation between the simulation density field
and the reconstructions, as expected from the contour plots of Figure 6.
The demonstration of these correlations is one of our major results.
Figures 6 and 7 reinforce the point made analytically in \S 2,
that non-gravitational models can exhibit the same sorts of velocity-density
correlations that are predicted by gravitational instability theories.
Section 2 based this argument primarily on linear theory, but here we show that
the result holds even for a realistically clustered, non-linear model.

\topinsert
\capskip
\vskip 0.4truein
\centerline{\bf FIGURE 7}
\vskip 0.4truein
\medskip
{\eightpoint
\capskip
\noindent {\bf Figure 7} ---
Scatterplots of the \velden reconstructions of the flat
explosion model.  Format is the same as Figure 5.  Results are similar
to those for gravitational instability models, though there are
differences of detail that reflect the greater prominence of large,
empty voids in the explosion model.
Scatterplots for the open ($\Omega_S=\Omega=0.17$) and dark ($\Omega_S=0.1,$
$\Omega=1$) explosion models are nearly identical, except for differences
in normalization.
}
\medskip
\textskip
\endinsert

At both the $1200\kms$ and $600\kms$ smoothing lengths,
the scatter in the reconstructed densities of the
explosion model is somewhat larger
than it is in the gravitational models, particularly at the lowest densities.
The non-linear Zel'dovich reconstruction removes most of the curvature in
the density correlations,
as expected based on the discussion in \S 2.3,
but it does not do much to reduce the scatter.
The increased scatter reveals itself in the values of the correlation
coefficient $r$, which are lower than those in Figure 5, differing
mostly at the $600\kms$ smoothing length.
Greater scatter
arises partly
from the more non-linear nature of the voids in the explosion model ---
they are larger and deeper than those in the
gravitational models --- and partly
from the lack of velocity tracers in the large, empty voids.
In an observational analysis, many of the points at
lower densities would be eliminated from these scatterplots or given little
weight because of the poor knowledge of the velocity field, \eg they would
be flagged by the ``$R_4$'' criterion of DBF.
Observational errors would also add to the scatter in the gravitational
reconstructions.  It seems unlikely, therefore, that the subtle differences
in scatter and curvature that can be seen between Figures 5 and 7 would
be discernible in a realistic case.

Density contour plots and scatterplots for the open and dark explosion
models are very similar to those shown for the flat model in Figures 6 and 7,
except for differences in normalization that reflect the lower velocities
in these low-$\Omega_S$ models.
Furthermore, in the dark models we obtain the same correlations,
up to a constant factor, regardless of whether we compare the
reconstructed density to the baryonic mass density or to the total
mass density; gravity pulls the dark matter into the
pre-existing baryon perturbations, so on large scales there is a simple,
linear bias between the baryon fluctuations and the total mass fluctuations.
Table 2b lists the values of $\beta$ inferred
from the explosion reconstructions,
along with the true $\beta$ values.
The inferred $\beta$'s are $\sim 0.7$ for the flat model,
where the true $\beta=1$, and $\sim 0.3$ for the open model,
where the true $\beta=0.17^{0.6}=0.34$.
For the $\Omega=1$, dark model we measure the ratio of rms galaxy
fluctuations to rms mass fluctuations directly on the $600\kms$ and
$1200\kms$ smoothing scales, obtaining $\beta_{true}=0.25-0.26.$
Alternatively, we could define $b$ by the ratio of galaxy fluctuations
to {\it baryonic} mass fluctuations, as we did in \S 2, and use $\Omega_b$
instead of $\Omega_{tot}$ in the definition of $\beta$.
Since galaxies trace the baryon distribution in the dark explosion model,
this definition yields $\beta_{true}=0.1^{0.6}/1=0.25$, the same value
as before.  The former definition of $\beta$
will behave more sensibly with time,
increasing as the dark matter catches up to the baryons, but at this
epoch we get the same answer from either method.
The inferred $\beta$ values for the dark model range from 0.20 to 0.23.

\topinsert
\centerline{Table 2b: Inferred $\beta$ values for Explosion Models}
\bigskip
\hrule\vskip0.1truecm\hrule
$$\vbox{\tabskip 1em plus 2em minus 5em
\halign to\hsize{ #\hfil & \hfil # & \hfil # \hfil &
\hfil # \hfil & \hfil # \hfil & \hfil # \hfil \cr
{\hfil Model\hfil} &$R_s$&$\beta_{true}$&$\beta_1$&$\beta_2$&$\beta_1$\cr
     &&       & {\hfil \velden \hfil}\span & {\hfil \denvel
\hfil} \cr
\noalign{\medskip\hrule\medskip}
Flat  & 1200 & 1.00 & 0.76 &  0.71  & 0.83 \cr
      &  600 & 1.00 & 0.72 &  0.69  & 0.82 \cr
Open  & 1200 & 0.34 & 0.30 &  0.28  & 0.33 \cr
      &  600 & 0.34 & 0.29 &  0.28  & 0.32 \cr
Dark  & 1200 & 0.25 & 0.21 &  0.20  & 0.23 \cr
      &  600 & 0.25 & 0.23 &  0.23  & 0.23 \cr
\noalign{\medskip\hrule\medskip}
}}$$
\smallskip
\endinsert

The first thing to note about Table 2b is that
the inferred values of $\beta$ are
in the range $0.2-1$,
and hence are not unlike those derived from observations (\eg
Dekel \etal~1993).
More remarkable is the fact that the inferred values of $\beta$
correspond quite closely to the true values;
only for the flat model do the fractional
errors exceed 20\%.  In the absence of gravity, the velocities produced
by explosions (or by other non-gravitational perturbations) would
decay kinematically in a few expansion times, once the non-gravitational
forces themselves turned off.  However, gravity does play a role
if $\Omega>0$, and the velocities only decay until they reach the
level that can be sustained by the most rapidly growing
gravitational mode.  If the velocities lie below those that
gravity can sustain, then gravitational accelerations raise them
to the growing-mode amplitude in roughly a Hubble time.
In either case, the inferred $\beta$ approaches the true value at
late times, \ie\ several expansion factors after the non-gravitational
effects become unimportant.
The gravitational phase of our flat model lasts only an
expansion factor of 2, and in this model the inferred $\beta$'s deviate
significantly from the correct value.  Our other models expand by
factors of 4 or 6 in the gravitational phase, and they yield more
accurate $\beta$ estimates.

The match between the inferred and true $\beta$'s in Table 2b suggests
that the velocities in most of our explosion models are in fact
sustained by gravity at the final epoch.  It may seem churlish to
describe a model as non-gravitational if the final velocity field
is dominated by gravitational accelerations.
However, the distinction is far more than semantic.
First, one can construct models in which the non-gravitational
perturbations produce most of the {\it displacements}, and hence most
of the spatial structure, even if the final velocities are sustained
mainly by gravity.  Second, the physical basis of the ``seed''
fluctuations is completely different in gravitational and
non-gravitational models.
Most non-gravitational theories locate the source of structure in
``mundane'' astrophysics after recombination, while most gravitational
instability models assume that the seeds of structure arise from
``exotic'' processes in the very early universe.

We should emphasize that the density-velocity correlations in Figure 7
do not arise simply because gravitational accelerations have come
to dominate over the original velocities --- on large scales these
gravitational accelerations are simply
proportional to the original velocities,
so they affect the velocity divergence field only by a constant overall
factor.  In fact,
velocity-to-density reconstructions of the explosion models at the
end of the hydrodynamic phase, before any $N$-body evolution, yield
similar correlations, and we showed in \S 2.1 that these correlations
would arise even in a model with no gravity at all.
The correlation between density and velocity divergence reflects the
physics of the continuity equation, not gravity {\it per se}.
We should also note that the agreement between true and inferred $\beta$
values in non-gravitational models
generally develops only after several expansion factors, and that the value
of ``several'' depends on the amplitude of the original velocities and
on the values of $\Omega$ and $b$.

\bigskip
\centerline{\bf 4.2 Recovering the Velocity Field}
\medskip

Having compared the density fields reconstructed from velocities to the
actual simulation density fields, we now turn our attention to
the reverse task of reconstructing velocity fields from
smoothed density fields.  We adopt the
linear theory prescription
$$
\vct{v}_{G1}=-\vct{\nabla}\Phi_{v,G1} ~, \qquad
	     \gradsq\Phi_{v,G1} = (H\beta)\del_G ~,
\eqno(\new)$$
where $\del_G$ is the smoothed galaxy density contrast
(\cf equation \eqref{5}a).
We first determine the
velocity field assuming $\beta=1$; the actual value of $\beta$ can then be
inferred by demanding that the rms value of the reconstructed velocity
field equal the rms value of the simulation velocity field.

\bigskip
\centerline{\bf 4.2.1 Gravitational Models}
\medskip

Figure 8 shows a slice through the smoothed velocity field of one of
the unbiased gravitational simulations and a corresponding slice from
the reconstructed velocity field.  The smoothing length is $R_s=600\kms$.
The location of the slice is the same as that of the slice through the
density field in Figure 4.  Arrows mark the projection of the smoothed
velocity onto the $x-y$ plane; multiplying the lengths of the arrows
by $100h$ yields velocities in $\kms$.
The reconstructed velocity field matches the simulation velocity field
accurately in both direction and amplitude.
At $1200\kms$ smoothing (not shown), the agreement between the two
velocity fields is even better.

\topinsert
\capskip
\vskip 0.4truein
\centerline{\bf FIGURE 8}
\vskip 0.4truein
\smallskip
{\eightpoint
\capskip
\noindent {\bf Figure 8} ---
Density-to-velocity reconstruction of an unbiased gravitational model.
The left panel shows a slice through the smoothed velocity field of
an unbiased CDM simulation, with a $600 \kms$ smoothing length.
Vectors indicate the smoothed velocity projected onto the $x-y$ plane.
The right panel shows the same slice through the velocity field that is
reconstructed from the smoothed density field using the linear
approximation $\vct{v}_{G1}$ of equation (29).
The reconstruction recovers the direction and amplitude of the
true velocity field.
}
\bigskip
\capskip
\vskip 0.4truein
\centerline{\bf FIGURE 9}
\vskip 0.4truein
\medskip
{\eightpoint
\capskip
\noindent {\bf Figure 9} ---
($a$)
Scatterplots of the \denvel reconstructions of unbiased gravitational models.
Each panel plots the three components of the reconstructed velocity field
against the three components of the true velocity field ($\vct{v}_G$)
for a set of 512 pixels uniformly spaced throughout the simulation volume.
Left-hand panels show a smoothing length of $1200 \kms$, right-hand panels
a smoothing length of $600 \kms$.
Velocities are normalized by the rms 1-d
value of the corresponding field; each panel lists these values in $\kms$.
Diagonal lines mark the $y=x$ relation expected for a perfect reconstruction.
The points follow this relation with small scatter.
($b$) Same as ($a$), but for biased gravitational models.
}
\textskip
\endinsert

The scatterplots of Figure 9a compare the true and reconstructed
velocity fields.  At each of 512 uniformly spaced pixels, we plot the
$x$-, $y$-, and $z$-components of the reconstructed velocity field
against the corresponding components of the simulation velocity field.
We normalize the velocities by the rms (1-dimensional)
value of the corresponding
velocity field; these rms values are listed in each panel.
These plots demonstrate close agreement between the true and reconstructed
velocity fields at both smoothing lengths.
Non-linear effects are clearly much smaller here than in the density
reconstructions considered above.
This difference is expected because
at a given smoothing length peculiar velocities are more strongly affected
by larger scale perturbations, which are closer to the linear regime.
In addition, the constraint that densities remain non-negative necessarily
imposes an asymmetry between the evolution of positive and negative
density fluctuations once density contrasts approach unity.
No such constraint applies to velocities; indeed, isotropy implies
that there must be no systematic differences
between positive and negative velocities.
As a result, non-linear evolution of velocity fields cannot
produce the sort of curvature
that appears in the density scatterplots; only anti-symmetric distortions
that treat positive and negative velocities identically are consistent
with isotropy.  A subtle distortion of this sort can be seen in the
right-hand panel of Figure 9a.

Figure 9b shows velocity scatterplots for the biased gravitational model.
Results are similar to those in Figure 9a: linear correlations with
modest scatter.
Column 6 of Table 2a lists the values of $\beta$ derived from
these reconstructions (and averaged over the three independent runs).
For unbiased models, the inferred $\beta$'s range from 0.87 to 0.92,
and for biased models they range from 0.46 to 0.48.
The inferred $\beta$ values are systematically low, but they are more
accurate than the values derived from the \velden reconstructions
discussed earlier.  This difference again reflects the fact that velocity
fields are more linear than density fields at the same smoothing scale.

Since we derive our reconstructed velocity fields from velocity potentials,
they are irrotational by construction.  The success of the reconstructions
demonstrates that the smoothed velocity fields of the simulations are
themselves
nearly irrotational.  We can quantify this result by decomposing each
simulation velocity field into a component that is curl-free (irrotational)
and a residual component that is divergence-free, and which therefore cannot
be derived from a scalar velocity potential.  The rms amplitude of the
divergence-free velocity component is typically smaller than the rms amplitude
of the curl-free component by about a factor of five.

\bigskip
\centerline{\bf 4.2.2 Explosion Models}
\medskip

Figure 10 shows slices through smoothed velocity fields of an explosion
simulation and the corresponding reconstruction.
The format is the same as that of Figure 8, and the smoothing length
is $R_s=600\kms.$
The location of the slices is the same as that of the density slices
in Figure 6.
The true and reconstructed velocity fields agree very well.

\topinsert
\capskip
\vskip 0.4truein
\centerline{\bf FIGURE 10}
\vskip 0.4truein
\medskip
{\eightpoint
\capskip
\noindent {\bf Figure 10} ---
Density-to-velocity reconstruction of a simulation of the
flat explosion model,
with a smoothing length of $600 \kms$, in the same format as Figure 8.
The reconstruction accurately recovers the true smoothed velocity
field of this non-gravitational model.
}
\medskip
\textskip
\endinsert

Figure 11 plots the reconstructed velocities against
the true velocities, using the same format as Figure 9.
At both smoothing lengths
we find linear correlations with relatively little scatter.
Scatterplots for the open and dark explosion models (not shown) are nearly
identical.  Column 6 of Table 2b lists the values of $\beta$ derived
from these reconstructions.  The inferred $\beta$'s are similar to those
found from the \velden reconstructions, and the remarks at the end of
\S 4.1.2 apply here as well.

\topinsert
\capskip
\vskip 0.4truein
\centerline{\bf FIGURE 11}
\vskip 0.4truein
\medskip
{\eightpoint
\capskip
\noindent {\bf Figure 11} ---
Scatterplots of the \denvel reconstructions of the flat explosion
model, in the same format as Figure 9.  Results are similar to those
for the gravitational instability model, though the scatter in the
reconstructed velocity fields is somewhat larger, probably because of the
large voids in the explosion model.
}
\medskip
\textskip
\endinsert

The success of the \denvel reconstructions indicates that the velocity
fields as traced by the galaxies in these
explosion simulations are nearly irrotational.
If we decompose the velocity fields into curl-free and divergence-free
components, we find that the rms amplitude of the curl-free component
is larger by about a factor of five for $R_s=1200\kms$ and four for
$R_s=600\kms$.  The explosion velocity fields thus have about the same
amount of vorticity as the velocity fields of our gravitational models.
This is an important result because it indicates that the POTENT method
should recover the correct 3-dimensional velocity field from radial velocity
data even for the explosion models considered here.  We address this
point more directly in \S 6.

\bigskip
\centerline{\bf 5. Models That Fail}
\bigskip

The analytic arguments in \S 2 and the numerical examples in \S 4
demonstrate that some non-gravitational models can produce
correlations between density and velocity fields that are the same as
those predicted by the standard gravitational instability scenario,
though they may sometimes yield incorrect estimates of $\beta$ if
analyzed in the standard fashion.
The presence of such correlations does not, therefore, provide
evidence exclusively for the gravitational instability hypothesis.
However, the success of the models considered in \S 4 does not imply
that {\it all} plausible models will pass the velocity-density tests.
In this section we discuss some models that fail these tests, in the
sense that they yield only weak correlations between the
galaxy density field and the galaxy velocity field.  We focus our attention
on two models in which
the physical processes that influence
galaxy formation yield a distribution where
the galaxy density and the galaxy flow are not
related to each other as prescribed by the continuity equation.
Galaxies in these models are biased with respect to the baryonic
mass in a non-trivial way, so
regions of high and low galaxy density are not necessarily
associated with converging and diverging flows, respectively.

We first consider a model in which peculiar velocities and the evolution
of the mass distribution {\it are} driven by gravitational instability,
but where the formation of galaxies is modulated by a
non-gravitational process.  Specifically, we examine the
scenario proposed by Babul \& White (1991), who suggested that
radiation from high-redshift quasars might suppress nearby galaxy formation,
giving rise to apparent voids in the
galaxy distribution that are not truly empty of matter.
Babul \& White proposed this effect as a possible mechanism for reconciling
the standard CDM model with observations like the APM angular correlation
function (Maddox \etal~1990) and the QDOT counts-in-cells
(Efstathiou \etal~1990; Saunders \etal~1991), which imply strong galaxy
clustering on large scales.
They noted, however, that the model might have difficulty explaining the
observed correspondence between galaxy density and velocity fields, and
we show here that this concern is justified.
In order to simulate this ``voided'' CDM model,
we start with the mass distribution
of the unbiased CDM simulations analyzed in \S 4, but we generate the catalog
of galaxy particles by first excising all particles lying inside randomly
placed spheres of radius $15h^{-1}\mpc$ (the quasar ``exclusion'' zones),
then randomly sampling the remaining particle distribution to
a density of $0.01h^3\mpc^{-3}$.  The number of
exclusion zones is chosen so that they occupy $60\%$ of the volume in total.
We ignore any effects that quasars might have on the true or inferred
(via Tully-Fisher, $D_n-\sigma$, etc.) velocity field of the galaxies.

\topinsert
\capskip
\vskip 0.4truein
\centerline{\bf FIGURE 12}
\vskip 0.4truein
\medskip
{\eightpoint
\capskip
\noindent {\bf Figure 12} ---
Velocity-to-density reconstruction of the ``voided'' CDM
model, for comparison to Figure 5.
This model's galaxy distribution is generated from an unbiased CDM
simulation by excising all particles lying inside randomly placed,
spherical, ``quasar exclusion zones'' of
radius $15\hmpc$.  Even though gravity creates the peculiar velocities in this
model, the \velden reconstructions fail because the evolution of the
galaxy distribution does not obey the continuity equation.
}
\medskip
\capskip
\vskip 0.4truein
\centerline{\bf FIGURE 13}
\vskip 0.4truein
\medskip
{\eightpoint
\capskip
\noindent {\bf Figure 13} ---
Density-to-velocity reconstruction of the voided CDM
model, for comparison to Figure 9.  The false voids in this model make
spurious contributions to the reconstructed velocity field, though weak
correlations between the true and reconstructed fields remain.
}
\medskip
\textskip
\endinsert

Figure 12 shows scatterplots of the \velden reconstructions of
the voided CDM model, using the quasi-linear method and smoothing
lengths of 600 and $1200\kms.$
Figure 13 shows scatterplots of the \denvel reconstructions of this model.
Although the plots display some correlations, the scatter
is much larger than in our earlier reconstructions of either
gravitational or explosion models (compare
Figure 12 to Figures 5 and 7; Figure 13 to Figures 9 and 11).
Values of the correlation coefficient confirm the visual impression
of increased scatter; for example, at $1200\kms$ the correlation
coefficient is $r=0.63$ in both reconstructions of voided CDM,
while for the CDM and explosion models analyzed earlier the
correlation coefficients at this smoothing length always exceed 0.9.
Residual correlations are present in the voided model because
the galaxy distribution satisfies
the continuity equation in some regions of the simulation volume.
Regions of high galaxy density are associated with genuine high matter
density, and therefore with converging velocity flows.
These high density regions are consequently recovered in the reconstructed
density maps, while convergent flows are recovered in the reconstructed
velocity maps.  Violation of the continuity equation,
and therefore the failure of the reconstruction scheme, is more
likely to occur in regions of low galaxy density, as such regions
do not necessarily represent
low matter density.  A high mass density
fluctuation located in a region of suppressed galaxy formation will not
appear in the galaxy map, but through its gravity it
will make its presence felt
in velocity space.  Consequently, a density map reconstructed from the
velocity field will show a high density region where none exists in the galaxy
map. Similarly, the ``voids'' in the galaxy map make spurious contributions to
the predicted velocity flow pattern.

\topinsert
\capskip
\vskip 0.4truein
\centerline{\bf FIGURE 14}
\vskip 0.4truein
\medskip
{\eightpoint
\capskip
\noindent {\bf Figure 14} ---
Comparison of the reconstructed density contrast to the {\it mass} density
contrast $\del_\rho$ of the voided CDM model.  The \velden technique
recovers the mass distribution of this model with reasonable
but not high accuracy.
}
\medskip
\textskip
\endinsert

Although it does not recover
the galaxy map accurately, the
\velden reconstruction of the ``voided'' CDM model yields a reasonable map of
the underlying {\it mass} distribution.
This is not surprising, since the galaxy velocity field
traces the mass velocity field, and
the mass distribution necessarily obeys the continuity equation.
Figure 14 plots the reconstructed density against the mass density, and the
correlations are much tighter than in Figure 12.  Nonetheless, there is more
scatter here than in the reconstruction of the standard CDM model (Figure 5)
because the velocity field is not well sampled
inside the voids,
so that the smoothed (galaxy-weighted) velocity field is afflicted by
``sampling-gradient bias'' (see DBF).

\topinsert
\capskip
\vskip 0.4truein
\centerline{\bf FIGURE 15}
\vskip 0.4truein
\medskip
{\eightpoint
\capskip
\noindent {\bf Figure 15} ---
Velocity-to-density reconstruction of an explosion model in which
shells do not completely fill space, for comparison to Figure 7.
We assume that all shells have the same radius, and we assume that
galaxies form only on the shells.
This model leaves false voids
in the regions between shells, which are empty of galaxies but not of mass.
The \velden reconstructions fail
drastically because the evolution of the galaxy distribution does not
obey the continuity equation.
}
\medskip
\capskip
\vskip 0.4truein
\centerline{\bf FIGURE 16}
\vskip 0.4truein
\medskip
{\eightpoint
\capskip
\noindent {\bf Figure 16} ---
Density-to-velocity reconstruction of the shell-galaxy explosion model,
for comparison to Figure 11.  Only weak correlations between the true
and reconstructed velocity fields exist for this model.
}
\medskip
\textskip
\endinsert

As our second model, we consider an explosion scenario in which the
explosion blast waves do not completely fill space.  We further assume
that the high gas densities of a blast wave shock are a necessary
prerequisite for galaxy formation, so that galaxies form only on shells
(see Ostriker \& Cowie 1981; Vishniac 1983).  The simulations are similar
to those described in \S 3.2, but instead of using a power-law distribution
of shell radii, we place 16 shells of equal radius at random locations in
the simulation cube.  We choose the shell radius so that the formal filling
factor is $ff=0.75$ at the end of the hydrodynamic phase, and we evolve for
an expansion factor of four in the gravitational phase, adopting
$\Omega_S=\Omega=1$.  We identify as galaxies only those particles that
were swept up onto shells before the end of the hydrodynamic phase.
The resulting galaxy distribution contains a combination of true voids
in the baryon distribution, created by the explosions, and false voids,
which contain undisturbed baryons at the mean density but no galaxies.
The evolution of this distribution violates the continuity equation,
in the sense that some of the voids are created by
suppressing galaxy formation instead of by moving galaxies out;
these voids do not have corresponding outflow signatures in the
peculiar velocity field.
Figures 15 and 16 show scatterplots of the
\velden and \denvel reconstructions, respectively, for this version
of the explosion model.  Some correlations are present, but
correlation coefficients are low, and
the scatter
is very large, much larger than it is for the model in which galaxies
faithfully trace the
baryonic
mass (\cf Figures 7 and 11).
Correlations arise in regions where shells have collided and eliminated
the false voids between them.  Consequently, we expect correlations in this
model to grow with time as the shells expand to fill all of space.
Qualitatively, results for this modified explosion model and results for
the voided CDM model are quite similar.
An important common feature is that the scatter
in the velocity-density correlations does not drop significantly
as one goes from $600\kms$ to $1200\kms$ smoothing length; in some cases
it even goes up.  This is in stark contrast to our results in \S 4,
where scatter arises from non-linear effects, and correlations are always
tighter at the larger smoothing length.

We have also analyzed the single-shell-radius explosion models assuming
that galaxies form with equal efficiency everywhere, so that they trace
the baryonic mass distribution.  In this case the model does obey the
continuity equation, and the velocity-density correlations are similar to those
obtained for the power-law explosion model studied in \S 4.
As mentioned in \S 3.2, it is natural to assume that galaxies trace the
baryonic mass in the power-law model,
since the explosion blast waves completely fill space.

We argued in \S 2 that a gravitational or non-gravitational model would pass
the velocity-density tests, at least on scales in the linear regime,
provided that (a) the galaxy distribution obeys the continuity equation,
in the sense that the distribution is initially uniform and fluctuations
are created only by moving galaxies from one place to another, and
(b) the present-day galaxy velocity field is irrotational and
proportional to the time-averaged
galaxy velocity field, and hence to galaxy displacements.
The biased CDM model of \S 4 actually violates condition
(a) because the galaxies are ``born clustered,'' but the final galaxy
fluctuations are approximately proportional to fluctuations in the
mass distribution, which does obey the continuity equation.
The two models that we have considered in this section fail
the velocity-density
tests because the processes that modulate galaxy formation
violate condition (a) in more radical ways, yielding
a non-trivial bias between the galaxy distribution and the
underlying distribution of baryonic matter.

One can also imagine plausible variations of the explosion model
that violate condition (b).
We have so far
considered explosion models in which all of the blasts are coeval:
the ratio of expansion velocity to Hubble velocity is the same for
every unperturbed shell.
If explosions occur over a wide range of epochs, on the other hand,
then the velocities associated with different blasts will
decay by different factors, and no single constant of proportionality
will relate present-day velocities to time-averaged velocities.  While
the relation $\div\vct{v}\propto -\delta$ will hold in local
regions, the constant of proportionality (and hence the inferred value
of $\Omega$) will vary from place to place, adding scatter to any global
correlations between the velocity and density fields.
However, this spatial variation will die out as gravitationally induced
velocities come to dominate over residual velocities from the explosions,
and for reasonable parameter choices it may well be unimportant.

Condition (b) is also violated to some degree
in the hybrid model discussed by
Thompson \& Park (1992), where explosions perturb the baryon distribution
but there are {\it independent} fluctuations in a gravitationally dominant,
collisionless dark matter component.  As the explosion velocities decay
and gravitational accelerations induced by the dark matter grow,
velocities of the baryon fluid will change in both amplitude and direction,
destroying the link between velocity and displacement.
This hybrid model behaves quite differently from the explosion models
with dark matter that we considered in \S 2 and \S 4; there the
dark matter perturbations develop {\it in response} to the explosion-induced
baryon perturbations, so they produce gravitational accelerations that are
parallel and proportional to the explosion velocities.
One would expect weaker velocity-density correlations in a hybrid model,
but quantitative
predictions are difficult without detailed modelling, since the baryons
and dark matter affect each other in ways that depend on $\Omega_b$ and
on the relative perturbation strengths.

\bigskip
\centerline{\bf 6. POTENT Reconstructions}
\bigskip

In sections 4 and 5, we reconstructed the density maps for various models
using the full 3-d velocity fields traced by the galaxies in the simulations.
However, real observations provide only the radial
component of peculiar velocities.
If the velocity field is a potential flow,
then one can recover the 3-d velocity field from the radial velocity
field using the POTENT method of Bertschinger \& Dekel (1989).
In this scheme, one integrates the radial velocity field along radial rays to
construct the velocity potential; the gradient of this potential yields
the 3-d velocity field.
The associated density field can be computed by the prescriptions
discussed in \S 2.

We have already commented that the vorticity of the smoothed velocity
fields in our models is small.  We expect, therefore, that the POTENT
procedure should derive fairly accurate 3-d velocity fields given
the radial velocity fields of these models, and that results for
\velden reconstructions should be similar to those obtained using the
full 3-d fields.  In this section, we demonstrate these points explicitly
by analyzing three of our simulations using the full POTENT machinery.

{}From each simulation, we create a data set containing the positions and
radial peculiar velocities of galaxies within $8,000 \kms$ of an
``observer'' at the center of the cube (using periodic replicas of the
simulation volume where necessary).
We do not introduce
errors in the positions and velocities of the ``galaxies.''
We analyze these simulated data
sets using the techniques developed by DBF for analysis of
observational data.
We smooth the discrete radial velocity data
onto a spherical grid using a tensor window with a
spherical Gaussian weighting function of radius $1000\kms$.
We also weight the contribution of each galaxy by the volume enclosing
it and its four nearest neighbors, in order to minimize the
``sampling-gradient bias'' that results
from inhomogeneous sampling of the radial velocity field.
The velocity potential at each spherical grid-point is computed
by integrating the smoothed radial velocity field
along rays originating from observer's position in the center of the
catalog volume.  We interpolate the velocity potential onto
a cubic grid with a spacing of $500\kms$ using a cloud-in-cell scheme,
and differentiate it to obtain the estimated 3-d velocity field (${\vec v}_P$)
in Cartesian coordinates.  The corresponding density field
$\delta_P$ is reconstructed using the quasi-linear procedure outlined in
\S 2.3.
For a more detailed exposition of the smoothing and weighting
scheme, we refer the reader to DBF.

\topinsert
\capskip
\vskip 0.4truein
\centerline{\bf FIGURE 17}
\vskip 0.4truein
\medskip
{\eightpoint
\capskip
\noindent {\bf Figure 17} ---
POTENT reconstructions of simulated data sets drawn from
three of our models.  The input data sets contain positions and radial
velocities of particles within $8,000 \kms$ of a central ``observer.''
The 3-d velocity field, smoothed with a Gaussian window of radius
$1,000 \kms$, is recovered by integrating the smoothed radial field
along radial rays and differentiating the resulting potential.
The density field is reconstructed from the divergence of the 3-d velocity
field by the quasi-linear method discussed in \S 2.3.
Left hand panels plot the smoothed, 3-d velocities recovered by POTENT
against the true velocities smoothed directly in 3-d.
Right hand panels plot the POTENT density contrast against the model galaxy
density contrast.  Reconstructions of the velocity and density fields
of the unbiased CDM model (top panels) and the explosion model
(middle panel) are quite successful because the velocity fields are
nearly irrotational and the galaxy distributions obey the continuity equation.
Recovered velocities of the shell-galaxy explosion model (lower left
panel) show more scatter because of the poor sampling of the velocity
field in this heavily voided model.  Density reconstruction of this
model (lower right panel) fails completely because the galaxy distribution
is biased in a way that violates the continuity equation; this result
is similar to that obtained directly from the 3-d velocity field
(Figure 15).
}
\medskip
\textskip
\endinsert

Figure 17 shows
scatterplots of the true and the
POTENT-reconstructed density and velocity fields for three simulations:
the unbiased CDM model (top panels), the flat explosion model
(middle panels), and the modified explosion model discussed in \S 5, where
galaxies form only on shells (bottom panels).
We compare the fields inside a sphere of radius $6000\kms$ instead of
the full $8000\kms$ volume used for the reconstruction, in order to avoid
edge effects.
We smooth the true density and velocity fields using a
$1000\kms$ Gaussian window, and to ensure a fair comparison we also
weight the true velocity field by the 4th-neighbor volume.

{}From the upper panels, we see that POTENT is quite successful at recovering
the velocity and density fields of the unbiased CDM simulation.
These results are unsurprising, since the velocity field is very nearly
a potential flow on these scales, and since the model satisfies the
other conditions required for successful \velden reconstructions.
The results from the full POTENT procedure are similar to those obtained
directly from 3-d velocity field at a smoothing length of $1200 \kms$
(lower left panel of Figure 5a), with a slightly larger scatter that
reflects both the shorter smoothing length and the imperfect recovery
of the 3-d velocity field in the POTENT analysis.

{}From the middle panels, we see that POTENT successfully recovers the
velocity and density fields of the explosion model.
There is some scatter in the recovered velocity field, presumably reflecting
the presence of a small amount of vorticity in the flow and
the absence of velocity tracers in the large voids, which
induces sampling-gradient bias.  Nonetheless, the plot of reconstructed
density versus true density is quite similar to that obtained directly
from the 3-d field at $1200 \kms$ smoothing (lower left panel of
Figure 7).  This result is important, as it reinforces our claim
that non-gravitational models, analyzed by POTENT, can yield the same
correlations between velocity and density that are found in standard
gravitational models, and in observations.

POTENT is less successful at recovering the velocity field of the
shell-galaxy explosion model --- the correlation coefficient between
reconstructed and true velocities is only $r=0.64$, compared to
$r=0.84$ and $r=0.87$ for the other two models.  We should emphasize
that this reconstruction of the 3-d velocity field has nothing to do
with the \denvel reconstructions discussed in previous sections;
the input is the radial velocity field, not the galaxy density field,
and the essential assumption is that of potential flow, not the
continuity equation.  The relatively poor recovery of the velocity
field reflects the dominance of large voids in the galaxy distribution.
Galaxies in this model provide poor spatial tracers of the velocity field,
so the smoothed radial velocity field suffers from sampling-gradient bias.
The reference ``true'' velocity field suffers from this bias as well,
and the bias may artificially induce vorticity that cannot be recovered
by POTENT (see the discussion at the beginning of \S 4).

While the velocity reconstruction in this model is less than perfect,
the failure of the density reconstruction is catastrophic,
yielding a correlation coefficient of only 0.24
(see the lower right panel of Figure 17).
This failure is unsurprising, and it reflects the fact that the galaxy
distribution in the shell-galaxy explosion
model does not evolve according to the continuity
equation.  The results are quite similar to those obtained directly
from the 3-d velocity field (Figure 15); even if POTENT recovered the
velocity field perfectly, the \velden analysis would not recover
the galaxy density field.  In application to observations,
comparisons of POTENT density fields to the density fields of IRAS or
optical galaxies typically yield correlation coefficients higher than 0.5,
so this model is probably excluded by existing data.

The POTENT analysis reinforces our previous findings
regarding \velden reconstructions.
POTENT densities should correlate with galaxy densities so long as
the galaxy distribution satisfies the continuity equation and the
velocity field is irrotational, whether or not the velocities are
gravitationally induced.  If physical processes modulate galaxy formation
in a way that violates continuity, then the reconstructed density field
will not match the observed galaxy distribution.

\bigskip
\centerline{\bf 7. Conclusions}
\bigskip

The gravitational instability hypothesis predicts specific correlations
between large-scale density and velocity fields.  We have argued that
finding these correlations does not provide confirmation of
the gravitational instability hypothesis because the same results arise
in a wider class of models.  Specifically, one expects to find
these correlations in any model where (a) structure in the galaxy
distribution
grows from homogeneous initial conditions
in a way that obeys the continuity equation, and
(b) the present-day velocity field is
irrotational and
proportional to the time-averaged
velocity field. The explosion models analyzed in \S 4 provide particular
examples of non-gravitational models that pass the usual density-velocity
tests.

In gravitational models, the ratio of the galaxy density contrast to the
divergence of the peculiar velocity field yields an estimate of
$\Omega$ (or of $\beta=\Omega^{0.6}/b$ in the case of linear biasing).
In non-gravitational models, the peculiar velocity field may contain a
mixture of growing and decaying modes, so the value of $\beta$ estimated
from this ratio may not correspond to the true cosmological value.
However, if non-gravitational forces shut off at some epoch, then modes
of the velocity field that are not sustained by gravitational
accelerations decay thereafter, on the cosmic expansion timescale.
As a result, one can find models in which non-gravitational forces initiate
the formation of spatial structure but gravity dominates the present-day
velocity field, and in such models the estimated $\beta$ should coincide
with the true value.

One can adopt either a pessimistic or an optimistic reading of
our results.
The pessimist would emphasize our conclusion that density-velocity comparisons
do not fundamentally test gravitational instability.  The value of $\beta$
estimated from these comparisons is guaranteed to
reflect the true cosmological value only if the gravitational hypothesis
holds, and the large-scale fields used for these estimates do not provide
the means to test this assumption.  Even agreement between the $\beta$
estimated from large-scale fields and values estimated by other techniques
would not imply that spatial structure formed by gravitational instability,
only that gravitational accelerations sustain the present-day velocities.

The optimist would note, first of all, that there are physically
interesting models that fail
the density-velocity tests.  These include models like voided CDM
(\cf \S 5 and Babul \& White 1991) in which the mass fluctuations and
peculiar velocities {\it are} generated by gravity but where galaxy formation
is modulated on large scales by non-gravitational processes,
which bias the galaxy distribution in a way that violates the continuity
equation.  Also included are models in which
hydrodynamic forces sculpt the spatial distribution of the baryons
but independent fluctuations in a collisionless dark matter component
produce gravitational accelerations that dominate the present
velocity field, since these independent accelerations break the link
between present and time-averaged velocities.  The optimist would also
note that in the broad class of models for which present and time-averaged
velocities {\it are} proportional, differentiating the velocity field does
yield a scaled map of the true mass distribution (with some unknown
scaling factor), even if the velocities have a non-gravitational origin,
and even if most of the mass is non-luminous.
Thus, comparisons of large-scale, galaxy density and velocity
fields offer a fairly general way to study
the relation between the distributions of galaxies and mass in the universe,
although they do not tell us whether gravity created these distributions.

\bigskip
\medskip

We thank Michael Strauss, Rien van de Weygaert, and Chris Thompson
for helpful comments on the manuscript, and Changbom Park for the use
of his cosmological $N$-body code.  AB and DHW acknowledge the support
of NATO postdoctoral fellowships at the Institute of Astronomy
(University of Cambridge), where this work began.  They also
acknowledge the hospitality of The Institute for Advanced Studies at
Hebrew University during the 1990 Jerusalem workshop on large-scale
structure and peculiar motions; AB further acknowledges the
hospitality of Tsvi Piran and the Racah Institute of Physics at Hebrew
University during the course of the workshop.
DHW acknowledges additional support from the W. M. Keck Foundation,
the Ambrose Monell Foundation, and NSF grant PHY92-45317 at the
Institute for Advanced Study.
AD acknowledges support from the US-Israel BSF Grant 89-00194.
JPO was supported in part by NASA grant NAGW-2448 and NSF grant AST91-09103.
He thanks the friends and colleagues who have listened to his expostulations
on this subject over the years; the resulting exchanges have helped to sharpen
the arguments given in this paper.

\vfill\eject

\def\refitem{\par\noindent\hangindent 20pt}
\parskip=3pt

\centerline{\bf References}
\bigskip

\refitem
Aaronson, M., Bothun, G., Mould, J., Huchra, J., Schommer, R. \& Cornell, M.
1986, ApJ, 302, 536

\refitem
Babul, A. \& White S.D.M. 1991, MNRAS, 253, 31p

\refitem
Bardeen, J.M., Bond, J.R., Kaiser, N. \& Szalay, A. 1986, ApJ, 304, 15


\refitem
Bertschinger, E. 1985, ApJS, 58, 1.

\refitem
Bertschinger, E. \& Dekel, A. 1989, ApJ, 336, L5

\refitem
Binney, J. 1984, MNRAS, 168, 73

\refitem
Burgers, J.M. 1974, The Nonlinear Diffusion Equation, Reidel, Dordrecht

\refitem
Burstein, D. 1990, Rep. Prog. Phys., 53, 421

\refitem
Cen, R. \& Ostriker, J.P. 1992, ApJ, 399, L113

\refitem
Davis, M. \& Peebles, P.J.E. 1983, ApJ, 267,465

\refitem
Davis, M., Strauss, M.A. \& Yahil, A. 1991, ApJ, 372, 394

\refitem
Dekel, A., Bertschinger, E. \& Faber. S.M. 1990, ApJ, 364, 349  (DBF)

\refitem
Dekel, A., Bertschinger, E., Yahil, A., Strauss, M.A., Davis, M. \& Huchra,
J.P. 1993, ApJ, in press

\refitem
Dekel, A. \& Rees, M.J. 1987, Nature, 326, 455

\refitem
Dressler, A., Lynden-Bell, D., Burstein, D., Davies, R.L., Faber, S.M.,
Terlevich, R.J. \& Wegner, G. 1987, ApJ, 313, 42

\refitem
Efstathiou, G., Kaiser, N., Saunders, W., Lawrence, A., Rowan-Robinson, M.,
Ellis, R.S. \& Frenk, C.S., 1990, MNRAS, 247, 10p

\refitem
Fisher, K. 1992, Ph.D. Thesis, U.C. Berkeley

\refitem
Gelb,  J.M. 1992, Ph.D. Thesis, M.I.T.


\refitem
Gurbatov, S.N., Saichev, A.I. \& Shandarin, S.F. 1985, Sov. Phys. Dokl., 30,
921

\refitem
Gurbatov, S.N., Saichev, A.I. \& Shandarin, S.F. 1989, MNRAS, 236, 385

\refitem
Harmon, R.T., Lahav, O. \& Meurs, E.J.A. 1987, MNRAS 228, 5p

\refitem
Hogan, C.J. \& Kaiser, N. 1983, ApJ, 274, 7

\refitem
Hogan, C.J. \& Kaiser, N. 1989, MNRAS, 237, 31p

\refitem
Hogan, C.J. \& White, S.D.M. 1986, Nature, 321, 575

\refitem
Hudson, M. \& Dekel, A. 1993, submitted to ApJ

\refitem
Ikeuchi, S. 1981, Pub. Astr. Soc. Japan, 33, 211

\refitem
Kaiser, N., Efstathiou, G., Ellis, R., Frenk, C., Lawrence, A.,
Rowan-Robinson,  M. \& Saunders, W. 1991, MNRAS, 252, 1

\refitem
Kaiser, N. \& Stebbins, A. 1991, Large-scale Structures and Peculiar Motions
in the Universe, ed. D.W. Latham and L.N. da Costa, Astronomical Society of
the Pacific, San Francisco, p111

\refitem
Katz, N., Hernquist, L. \& Weinberg, D.H. 1992, ApJ, 399, L109

\refitem
Lahav, O. 1987, MNRAS, 225, 213

\refitem
Lahav, O., Rowan-Robinson, M. \& Lynden-Bell, D. 1988, MNRAS, 234, 677

\refitem
Lubin, P. \& Villela, T. 1986, Galaxy Distances and Deviations from the
Universal Expansion, ed. B.F. Madore \& R.B. Tully, Reidel, Dordrecht, p169

\refitem
Lynden-Bell, D., Lahav, O. \& Burstein, D. 1989, MNRAS 234, 677

\refitem
Maddox, S.J., Efstathiou, G., Sutherland, W.J. \& Loveday, J. 1990, MNRAS,
242, 43p

\refitem
Meiksin, A. \& Davis, M. 1986, AJ, 91, 191

\refitem
Melott, A.L. 1986, Phys. Rev. Lett., 56, 1992

\refitem
Nusser, A., Dekel, A., Bertschinger, E. \& Blumenthal, G.R. 1991, ApJ, 379, 6

\refitem
Olson, D.W. \& Sachs, R.K. 1973, ApJ, 185, 91

\refitem
Ostriker, J.P. 1993, ARAA, 31, in press

\refitem
Ostriker, J.P. \& Cowie, L.L. 1981, ApJ, 243, L127

\refitem
Ostriker, J.P. \& Strassler, M. 1989, ApJ, 338, 579

\refitem
Ostriker, J.P., Thompson, C. \& Witten, E. 1986, Phys. Lett. B180, 231

\refitem
Park, C. 1990, MNRAS, 242, 59p (also: Ph.D. Thesis, Princeton University)

\refitem
Park, C. 1991, MNRAS, 251, 167

\refitem
Peebles, P.J.E. 1980, The Large-Scale Structure of the Universe,
Princeton University Press, Princeton

\refitem
Press, W.H., Teukolsky, S.A., Vetterling, W.T. \& Flannery, B.P. 1992,
Numerical Recipes, Cambridge University Press, Cambridge

\refitem
Rowan-Robinson \etal\ 1990, MNRAS, 247, 1

\refitem
Saarinen, S., Dekel, A. \& Carr, B.J. 1987, Nature, 325, 598

\refitem
Saunders, W., Frenk, C.S., Rowan-Robinson, M., Efstathiou, G., Lawrence, A.,
Kaiser, N., Ellis, R.S., Crawford, J., Xiao-Yang, X. \& Parry, I. 1991,
Nature, 349, 32

\refitem
Scaramella, R., Vettolani, G. \& Zamorani, G. 1991, ApJ, 376, L1

\refitem
Smoot, G.F., Gorenstein, M.V. \& Muller, R.A. 1977, Phys. Rev. Lett., 39, 898

\refitem
Smoot, G.F. \& Lubin, P.M. 1979, ApJ, 234, L83

\refitem
Strauss, M.A., Yahil, A., Davis, M., Huchra, J.P. \& Fisher, K.B.
1992a, ApJ, 397, 395

\refitem
Strauss, M.A., Huchra, J.P., Davis, M., Yahil, A., Fisher, K.B.
\& Tonry, J.P. 1992b, ApJS, 83, 29

\refitem
Strauss, M.A. \& Davis. M. 1991, Large-Scale Structures and Peculiar Motions
in the Universe: ASP Conf. Series No. 15, ed. D.W. Latham \& L.N. da Costa,
Astronomical Society of the Pacific, San Francisco, p53

\refitem
Thompson, C. \& Park, C. 1992, preprint

\refitem
Vishniac, E. 1983, ApJ, 274, 152

\refitem
Weinberg, D.H. \& Cole, S. 1992, MNRAS, 259, 652

\refitem
Weinberg, D.H. \& Gunn, J.E. 1990, MNRAS, 247, 260

\refitem
Weinberg, D.H., Ostriker, J.P. \& Dekel, A. 1989, ApJ, 336, 9

\refitem
White, S.D.M., Davis, M., Efstathiou, G. \& Frenk, C.S. 1987, Nature 330, 451

\refitem
Yahil, A. 1988, Large-Scale Motions in the Universe: A Vatican Study Week,
ed. V.C. Rubin and G.V.  Coyle, Princeton University Press, Princeton, p219

\refitem
Zel'dovich, Ya. B. 1970, A\&A, 5, 84

\bye